\documentclass[journal]{IEEEtran}
\usepackage{ifpdf}

\usepackage{cite}

\ifCLASSINFOpdf
  \usepackage[pdftex]{graphicx}
\else
  \usepackage[dvips]{graphicx}
\fi
\usepackage{amsmath, amssymb}
\usepackage{algorithmic, algorithm}

\usepackage{array}

\ifCLASSOPTIONcompsoc
 \usepackage[caption=false,font=normalsize,labelfont=sf,textfont=sf]{subfig}
\else
 \usepackage[caption=false,font=footnotesize]{subfig}
\fi

\usepackage{stfloats}
\ifCLASSOPTIONcaptionsoff
 \usepackage[nomarkers]{endfloat}
\let\MYoriglatexcaption\caption
\renewcommand{\caption}[2][\relax]{\MYoriglatexcaption[#2]{#2}}
\fi
\usepackage{url}

\usepackage[colorlinks,linkcolor=red,anchorcolor=green,citecolor=blue]{hyperref}

\usepackage{multirow}
\usepackage[table,xcdraw]{xcolor}
\usepackage{longtable}
\usepackage{subfig}
\usepackage{booktabs}
\usepackage{threeparttable}
\usepackage{lscape}

\begin{document}

\newcommand{\AJ}[1]{\textcolor{magenta}{[AJ: #1]}}
\newcommand{\FC}[1]{\textcolor{blue}{#1}}
\newcommand{\RoundOne}[1]{\textcolor{black}{#1}}

%
\title{An Algorithm-Hardware Co-Optimized Framework for Accelerating \textit{N:M} Sparse Transformers}
%
%

%

\author{Chao~Fang,~\IEEEmembership{Graduate~Student~Member,~IEEE,}
        Aojun~Zhou, 
        and~Zhongfeng~Wang,~\IEEEmembership{Fellow,~IEEE}
\thanks{This work was supported in part by the National Natural Science Foundation of China under Grant 62174084, 62104097 and in part by the High-Level Personnel Project of Jiangsu Province under Grant JSSCBS20210034, the Key Research Plan of Jiangsu Province of China under Grant BE2019003-4. \textit{(Corresponding author: Zhongfeng Wang.)}}
\thanks{C. Fang and Z. Wang are with the School of Electronic Science and Engineering, Nanjing University, Nanjing 210008, China (e-mail: fantasysee@smail.nju.edu.cn;
zfwang@nju.edu.cn).}
\thanks{A. Zhou is with CUHK-Sensetime Joint Lab, CUHK, Hong Kong, China (e-mail: aojunzhou@gmail.com).}
}

%
%

\markboth{Journal of \LaTeX\ Class Files,~Vol.~14, No.~8, August~2015}%
{Shell \MakeLowercase{\textit{et al.}}: Bare Demo of IEEEtran.cls for IEEE Journals}
%



\maketitle

\begin{abstract}
The Transformer has been an indispensable staple in deep learning. 
However, for real-life applications, it is very challenging to deploy efficient Transformers due to immense parameters and operations of models.
To relieve this burden, exploiting sparsity is an effective approach to accelerate Transformers.
Newly emerging Ampere GPUs leverage a \textit{2:4} sparsity pattern to achieve model acceleration, while it can hardly meet the diverse algorithm and hardware constraints when deploying models.
By contrast, we propose an algorithm-hardware co-optimized framework to flexibly and efficiently accelerate Transformers by utilizing general \textit{N:M} sparsity patterns.
(1) From algorithm perspective, we propose a sparsity inheritance mechanism along with inherited dynamic pruning (IDP) to obtain a series of \textit{N:M} sparse candidate Transformers rapidly.
A model compression scheme is further proposed to significantly reduce the storage requirement for deployment.
(2) From hardware perspective, we present a flexible and efficient hardware architecture, namely STA, to achieve significant speedup when deploying \textit{N:M} sparse Transformers.
STA features not only a computing engine unifying both sparse-dense and dense-dense matrix multiplications with high computational efficiency but also a scalable softmax module eliminating the latency from intermediate off-chip data communication.
Experimental results show that compared to other methods, \textit{N:M} sparse Transformers, generated using IDP, achieves an average of $6.7\%$ improvement on accuracy with high training efficiency.
Moreover, STA can achieve $14.47\times$ and $11.33\times$ speedup compared to Intel i9-9900X and NVIDIA RTX 2080 Ti, respectively, and perform $ 2.00 \sim 19.47 \times $  faster inference than the state-of-the-art FPGA-based accelerators for Transformers.
\end{abstract}


%
\IEEEpeerreviewmaketitle

\section{Introduction}

\IEEEPARstart{T}{ransformer-based} networks are a formidable force in deep learning \cite{tay2020efficient}.
Tremendous impact in many fields, such as neural machine translation \cite{song2020alignment}, language understanding \cite{kenton2019bert}, and image processing \cite{dosovitskiy2020image}, has been made since the innovation of Transformers. 
Nevertheless, the impressive performance of Transformers comes with heavy computing and memory costs, which become a significant barrier to the efficient deployment of Transformer-based applications.
Notably, BERT, a representative Transformer-based model \cite{kenton2019bert},  requires \textit{440MB} memory and over \textit{176G} floating-point operations. 
Such severe requirements on memory and computation make it critical to find an efficient solution for deploying Transformers.

\begin{figure} [htbp] 
	\centering
	\includegraphics[width=0.42\textwidth]{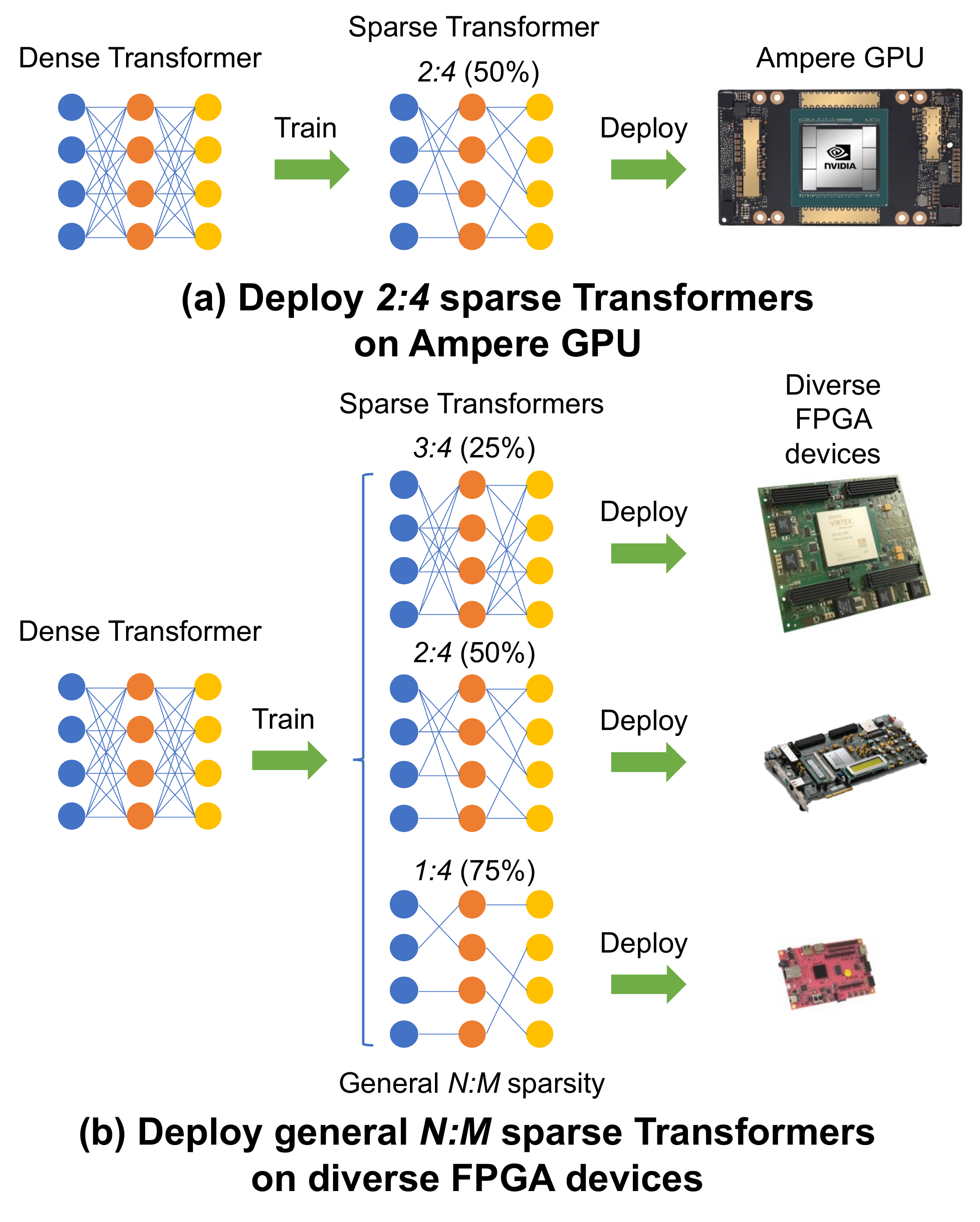}
	\caption{\RoundOne{Accelerating \textit{\textit{N:M}} sparse Transformer-based models (a) using modern Ampere GPUs and (b) using diverse FPGAs with our framework. 
	Compared to (a), (b) can generate a series of \textit{\textit{N:M}} sparse Transformers along with the dedicated accelerators for efficient model deployment.}
	} 
	\label{fig:sol_cmp}
\end{figure}

Sparsity is an important feature that can be utilized to improve the efficiency of DNNs deployment in dedicated accelerators.
In the pioneering works, OPTIMUS \cite{park2020optimus} and EdgeBERT \cite{tambe2021edgebert}, the latest ASIC accelerators, leverage unstructured sparsity to realize efficient deployment for Transformers.
Nevertheless, it is hard to predict the unstructured sparsity in advance, and therefore the acceleration performance can be greatly dragged.
Recent studies \cite{zhou2021learning} demonstrate deep neural networks leveraging \textit{N:M} fine-grained structured sparsity, where \textit{N} out of \textit{M} parameters are zeros for every continuous \textit{M} parameters, can achieve comparable performance over those leveraging unstructured sparsity \cite{sun2021dominosearch}. 
However, it is significantly restricted to accelerate \textit{N:M} sparse networks on current hardware platforms.
As shown in Fig.~\ref{fig:sol_cmp}~(a), the only existing solution is \textit{ASP with Ampere GPUs} that focuses on the middle-level (\textit{2:4}), i.e., 50\%, sparse ratio. 
Based on our experiments, a heavy Transformer can be dramatically slimmed by weight pruning with the aggressive \textit{N:M} pattern, e.g., \textit{2:8} or \textit{1:8}, achieving a considerable reduction in the amount of both parameters and operations.
The only choice of uniform \textit{2:4} sparsity limits performance when deploying Transformer-based models, making it inflexible to meet different hardware constraints (e.g., latency, energy).
Compared with the uniform \textit{2:4} sparsity, the more flexible general \textit{N:M} sparsity in real applications can satisfy various algorithm and hardware constraints under different deployment scenarios.
However, there is currently a lack of an integrated framework to investigate the deployment of Transformer with general \textit{N:M} sparse patterns. 
To bridge this gap, as presented in Fig.~\ref{fig:sol_cmp}~(b), we propose an algorithm-hardware co-optimized framework for accelerating \textit{N:M} sparse Transformers, which addresses two significant issues: \textbf{1). how to produce a series of \textit{N:M} sparse Transformers in an efficient way; 2). how to design a flexible and efficient dedicated architecture for \textit{N:M} sparse Transformers on diverse FPGA platforms.}
Although advanced optimization algorithms ASP\cite{mishra2021accelerating} and SR-STE\cite{zhou2021learning} can maintain the middle-level (\textit{2:4}) sparsity via static and dynamic fine-tuning, we observe existing methods degrade the performance significantly under high sparse ratio (e.g., $\geq 75\%$). 
In addition, the ASP and SR-STE schemes leverage single-shot magnitude-based pruning for a specified hyper-parameter \textit{N} and \textit{M}.  
This traditional recipe results in a significant performance drop with a higher sparse ratio and restricts the deployment of flexible \textit{N:M} sparse models on the FPGA platform. 
To overcome the aforementioned problems, we propose a sparsity inheritance mechanism, which increases the sparsity progressively to enable efficient searching for $N$:$M$ sparse Transformers under various sparsity configurations (e.g., \textit{2:8}, \textit{1:8}). 
We also propose a pruning method, namely inherited dynamic pruning (IDP), which shrinks pre-pruning models progressively, and the convergent pre-pruning initialized models can aid in the convergence of the following sub-networks. 
Extensive experiments are conducted on Transformer-based models, showing that models generated by IDP with the sparsity inheritance mechanism have superior performance on various sparsity ratios than those using the ASP and SR-STE.
Moreover, for efficient model deployment, we apply a simple but effective bitmap-based compression scheme, which dramatically reduce the storage requirements for \textit{N:M} sparse Transformers.


To enable flexible and efficient deployment on various FPGA devices, we design a highly configurable dedicated accelerator for \textit{N:M} sparse Transformers, namely STA.
STA fully explores the parallelism of Transformers in three aspects, including head parallelism, row parallelism, and column parallelism, which significantly improves computational efficiency.
It features two computing cores, a diverse matrix multiplication (MatMul) engine, called DMME, and a scalable softmax module, both of which are highly configurable.
Operations of \textit{N:M} sparse Transformers are dominated by two types of MatMuls.
One is the sparse-dense MatMul with \textit{N:M} sparse network parameters, and the other is dense-dense MatMul free of parameters.
DMME performs both sparse-dense and dense-dense MatMuls on-the-fly, and achieves much higher computational efficiency over the prior work \cite{lu2020hardware} under both modes.
Especially for sparse-dense MatMul, DMME only performs operations related to those remaining non-zero parameters, which greatly improves computational efficiency.
The scalable softmax module can perform the softmax function in Transformers. 
It keeps all the intermediate results fully local, eliminating latency from intermediate off-chip data communication.
According to the given architectural settings, STA can be rapidly implemented on FPGAs to realize efficient deployment for specific \textit{N:M} sparse Transformers.

To summarize, the contributions of our paper are as follows:
\begin{itemize}
    \item To our best knowledge, this is the first work that presents an algorithm-hardware co-optimized framework to systematically study the efficiency of fine-grained \textit{N:M} sparse Transformers on FPGA. The proposed framework can adjust to diverse hardware constraints for flexible and efficient model deployment.
    \item To generate a series of \textit{N:M} sparse Transformers simultaneously, we propose a sparsity inheritance mechanism along with the inherited dynamic pruning (IDP) algorithm, which can significantly achieve about $6.7\%$ accuracy improvement of Transformers under high sparsity compared with state-of-the-art methods.
    \item We present a simple but effective bitmap-based compression scheme for \textit{N:M} sparse Transformers compared to multiple sparse indexing formats, which dramatically reduces the storage requirements up to $5.33 \times$.
    \item We propose a dedicated hardware architecture, namely STA, to realize flexible and efficient deployment of \textit{N:M} sparse Transformers. It features two novel hardware modules handling intensive operations of Transformers, including a diverse matrix multiplication engine (DMME) that unifies dense and sparse MatMul operations in high computational efficiency, and a scalable softmax module to avoid frequent off-chip memory accesses.
    \item Extensive experiments have been conducted on four NLP tasks and four Transformer-based models to evaluate the effectiveness of the proposed framework, which achieves up to $19.47\times$ speedup over Intel i9-9900X, NVIDIA RTX 2080 Ti, and prior FPGA-based accelerators for Transformers.
\end{itemize}

The rest of this paper is organized as follows: Section~\ref{sec:bkg} presents an overview of Transformers, and state-of-the-art works for accelerating Transformers with innovations on hardware architecture. 
Section~\ref{sec:overview} introduces the workflow of our proposed algorithm-hardware co-optimization framework.
Section~\ref{sec:algo} and Section~\ref{sec:hwarch} elaborate optimizations on pruning algorithm and hardware architecture, respectively.
Comprehensive experimental results are presented in Section~\ref{sec:res} to show significant potential of our proposed co-optimization framework in Transformer-based applications.
\section{Background and Motivation} \label{sec:bkg}

In this section, we provide an overview of key structures in Transformers, and review related work on hardware accelerators for Transformers.


\subsection{Transformer Overview}

\begin{figure} [htbp] 
	\centering
	\includegraphics[width=0.48\textwidth]{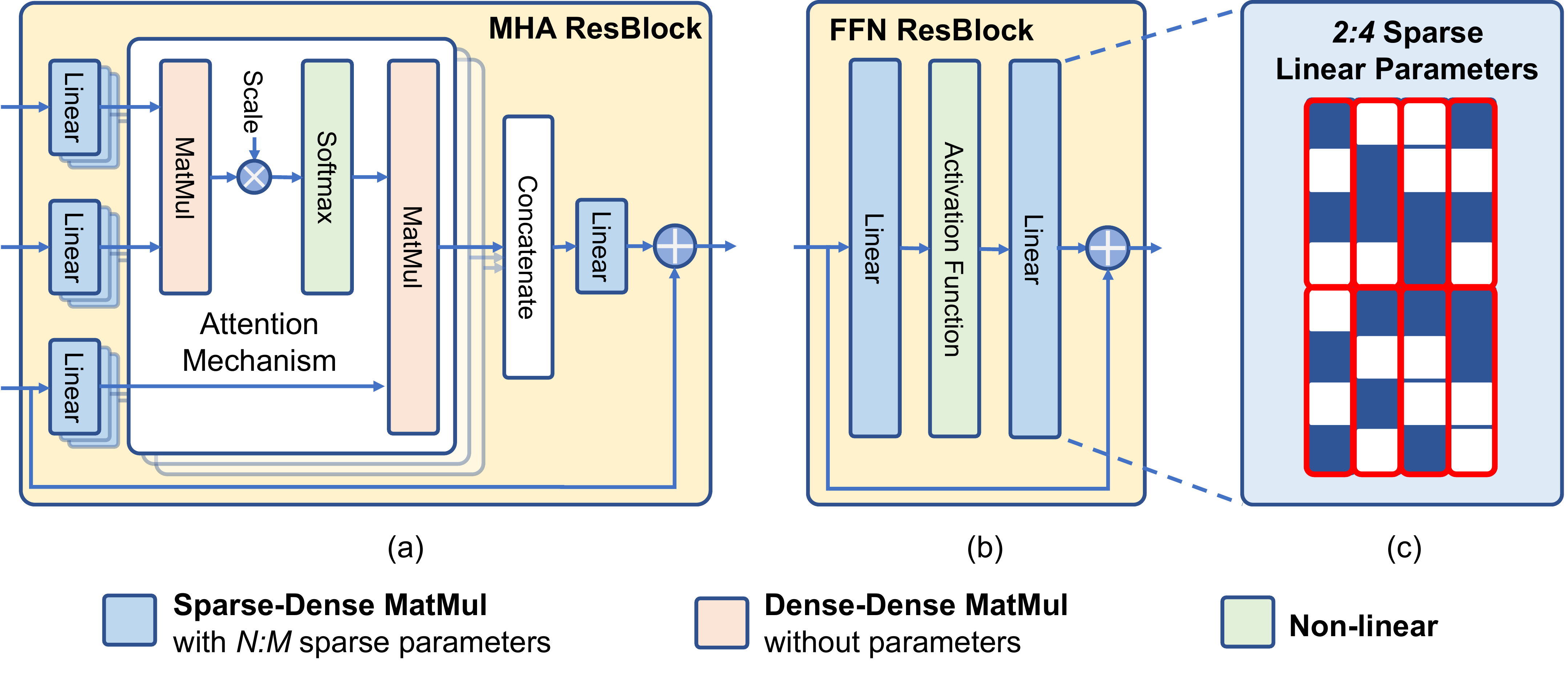}
	\caption{The operations in (a) the MHA ResBlock and (b) the FFN ResBlock under \textit{\textit{N:M}} sparsity pattern. Both Resblocks are the key structures of the Transformer. (c) An illustration of \textit{2:4} sparse parameters in a linear layer.} 
	\label{fig:model_overview}
\end{figure}

The key architectures of the Transformer \cite{vaswani2017attention} are characterized by a multi-head attention (MHA) residual block (ResBlock) and a position-wise feed-forward network (FFN) ResBlock.
Fig.~\ref{fig:model_overview}~(a) and (b) illustrate the inner structures of MHA and FFN ResBlocks, respectively.
The input and output of the FFN ResBlock are connected by a residual connector. 
And two linear transformation modules along with a activation function are inside the FFN ResBlock. 
The structure of MHA ResBlock is more complicated. 
The inputs of MHA ResBlock are split into multiple parallel heads with corresponding linear projection at first. 
Then the results are as input fed into the attention mechanism in parallel, and finally, the results of attention heads are concatenated together and passed into a
linear layer to obtain the output linear projection. 
Note that the attention mechanism is totally different from the linear layer,
performing parameter-free MatMuls. 
Thus, the computing engine for Transformers is required to support both sparse and dense MatMuls even though sparsity is introduced to parameters.
The residual connector of FFN ResBlock is organized the same as the FFN ResBlock.

\subsection{Recent Advances for Transformer Acceleration}

Extensive research has concentrated on the design of high-performance and energy-efficient DNN hardware accelerators \cite{wu2021swm, colleman2021high, yantir2021imca, yin2019vesti, paulin2021rnn, fang2021accelerating, yu2020uni, zhu2020efficient, moreno2020analysis, lian2019high, kala2019high, xie2021efficient, albericio2016cnvlutin, chen2016eyeriss, liu2016cambricon, liu2020systolic, parashar2017scnn}.
However, most of these works focus on CNN and RNN computations, and not as much scrutiny has been given to accelerating Transformer-based networks with self-attention mechanisms.

As a pioneer work, \cite{lu2020hardware} proposed a dense systolic array accelerator along with the partitioning scheme for FPGA-based acceleration of Transformers.
Moreover, FTRANS \cite{li2020ftrans} exploited block-circulant matrix-based weight representation for Transformer acceleration. 
However, both of them fail to utilize the sparsity of parameters in Transformers, thereby limiting the speedup of model deployment. 
A$^{3}$ \cite{ham2020a3}, SpAtten \cite{wang2021spatten}, and Sanger \cite{lu2021sanger} merely focused on the speedup potential for the sparse attention mechanism, all of which can hardly satisfy the needs of agile and efficient deployment of Transformer models. 
OPTIMUS \cite{park2020optimus} and EdgeBERT \cite{tambe2021edgebert} holistically accelerate Transformers with unstructured sparse matrix multiplications and save energy by skipping the computations related to those zero-value parameters.
Nevertheless, the unstructured sparsity leads to irregular data access, making both designs suffer low computational efficiency.
\cite{peng2021accelerating} exploited the coarse-grained block-based sparsity pattern for accelerating Transformers, while this sparsity pattern is so coarse that the models can hardly achieve a considerable sparsity ratio with acceptable accuracy. 



In summary, it is hard for all these works to achieve satisfying speedup and efficiency of Transformer deployment due to the lack of attention to model sparsity, limited sparse potential exploration on the whole Transformer models, or restricted computational efficiency for sparse Transformers.
To address above issues, this work presents an algorithm-hardware co-optimization framework to realize flexible and efficient deployment of Transformers by leveraging general \textit{N:M} sparsity patterns.
For algorithm optimization, we focus on how to generate a series of \textit{N:M} sparse Transformers in high quality and efficiency.
For hardware optimization, we concentrate on designing a flexible and efficient dedicated architecture that can accelerate \textit{N:M} sparse Transformers with high computational efficiency.
\begin{figure*} [ht] 
	\centering
	\includegraphics[width=0.92\textwidth]{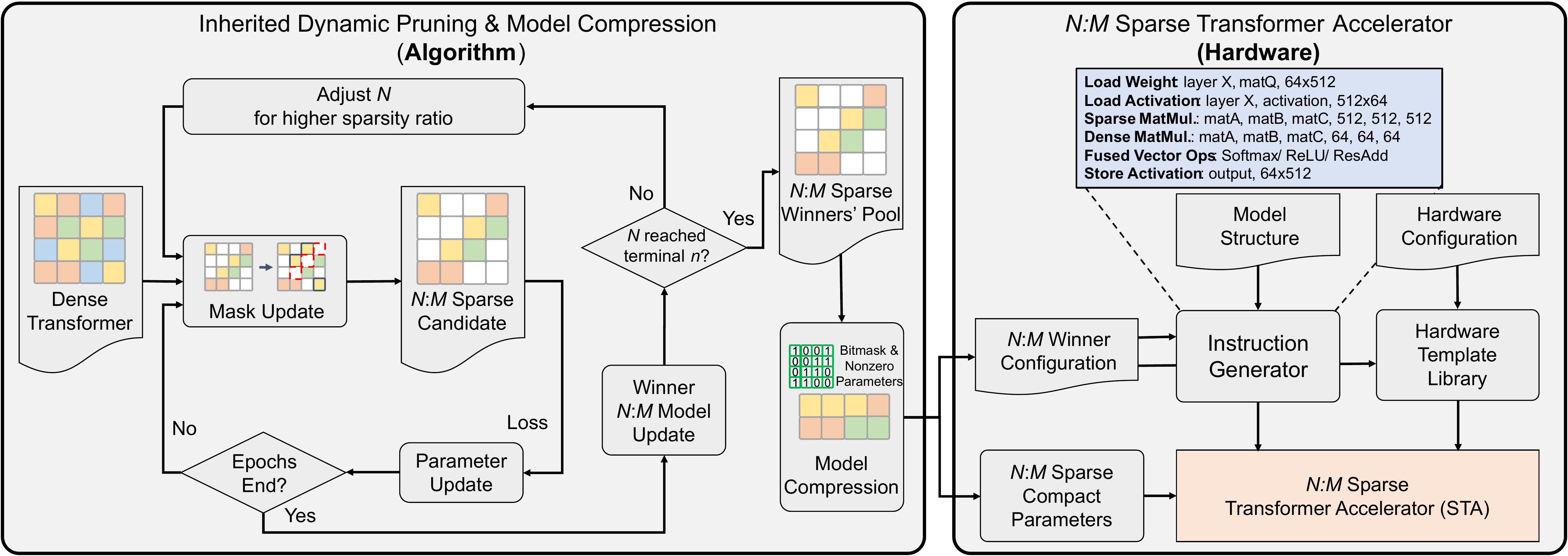}
	\caption{\RoundOne{The workflow of our proposed algorithm-hardware co-optimized framework. At the algorithm level, \textit{N:M} sparse Transformers can be rapidly generated by inherited dynamic pruning, and significantly compressed for further deployment. At the hardware level, the dedicated accelerator, STA, is implemented on the FPGA platform to accelerate the deployed \textit{N:M} sparse Transformer.}}
	\label{fig:workflow}
\end{figure*}

\section{Overview of Co-optimization} \label{sec:overview}


%
To achieve agile and efficient deployment of Transformers, we propose an algorithm-hardware co-optimized framework.
The overview of our framework is presented in Fig.~\ref{fig:workflow}.
According to the given specific requirements, our framework can quickly obtain the required \textit{N:M} sparse Transformer model with high accuracy, and provide corresponding Transformer accelerators on FPGA devices to realize efficient model deployment.
In this section, we elaborate on the workflow of our algorithm-hardware co-optimized framework.

At the algorithm level, we focus on quickly obtaining any desired \textit{N:M} sparse Transformer model, and achieving effective compression of the \textit{N:M} sparse Transformer.
The algorithm optimization is divided into two stages.
As shown in Fig.~\ref{fig:workflow}, the first stage is IDP based on the sparsity inheritance mechanism.
Compared with single-shot training \cite{mishra2021accelerating, zhou2021learning}, our method can utilize the knowledge of the previous \textit{N:M} sparse model, which contributes to faster and better convergence.
The second stage is model compression.
Only the non-zero parameters in the \textit{N:M} sparse Transformer would be stored, along with an additional binary mask that indicates position of all recorded elements.
The methods of pruning and model compression are presented in Sec.~\ref{sec:algo}.


At the hardware level, we concentrate on efficient and flexible hardware architecture design that boosts the computational efficiency for \textit{N:M} sparse Transformers.
The hardware optimization features an efficient hardware architecture for \textit{N:M} sparse Transformers along with an automatic hardware generator, which can meet requirements on various Transformer models, FPGA devices, and \textit{N:M} sparsity.
The automatic hardware generator is composed of instruction generator and hardware template library.
According to the \textit{N:M} configuration of the winner model and the network structure of the deployment model, the instruction generator can automatically produce instructions that guide STA to perform operations of the winner \textit{N:M} sparse Transformer.
As shown in Fig~\ref{fig:workflow}, instructions are divided into three categories: load/store data, sparse/dense MatMul operators, and fused vector operators.
The hardware template library can quickly generate a dedicated STA based on the pre-defined hardware configurations and the \textit{N:M} configuration of the winner model.
STA performs inference tasks for the Transformer model by accessing the compact sparse parameters and the pre-generated instructions.
STA, whose hardware architecture is elaborated in Sec.~\ref{sec:hwarch}, can achieve significant improvement on computational efficiency by eliminating all zero-valued parameter operations.

\RoundOne{
In real-life deployment, the choice of \textit{N:M} may change if there are multiple FPGA devices with different hardware resources, and varying deployment constraints including latency and model accuracy.
However, considering all the above factors, once an \textit{N:M} model is determined to be deployed, the model can meet the needs of practical applications.
Therefore, the \textit{N:M} would change before deploying models 
while would not change after the model deployment.
Compared to the Ampere GPU dedicated for \textit{2:4} sparse acceleration, specific \textit{N:M} STA can be flexibly configured and automatically generated on the selected FPGA device with significant performance gains benefited from dedicated \textit{N:M} sparse acceleration.
As \textit{N:M} changes, our framework would efficiently benefit from the algorithm-hardware co-optimization.
At the algorithm level, the proposed IDP could provide a series of \textit{N:M} models with varying computing complexity and model accuracy, among which we could select the most suitable one for further model deployment.
At the hardware level, the proposed STA could be flexibly generated based on the selected \textit{N:M} and other configurations, achieving significant acceleration of \textit{N:M} Transformers.
}

\section{Algorithm Optimization} \label{sec:algo}

In this section, we elaborated on algorithm optimizations of our framework.
Firstly, we demonstrate advantages of \textit{N:M} sparsity pattern in Sec.~\ref{subsec:pattern} by comparing it with other popular sparsity patterns. 
Then, pruning algorithm and compression scheme of \textit{N:M} sparse Transformers are presented in Sec.~\ref{subsec:pruning} and Sec.~\ref{subsec:packing}, respectively.



\subsection{\textit{N:M} Sparsity Pattern} \label{subsec:pattern}
A dense parameter matrix, can be pruned with a sparsity ratio of 50\% using three existing sparsity patterns, unstructured sparsity \cite{park2020optimus
}, block-based structured sparsity \cite{peng2021accelerating
}, and \textit{N:M} group-based structured sparsity \cite{zhou2021learning
}, respectively.
Table~\ref{tab:sparsity} summarizes all these pruning patterns.
Elements in any position of the parameter matrix can be pruned if the unstructured sparsity pattern is employed.
The unstructured sparse model can achieve a considerable compression ratio while maintaining comparable accuracy to the dense model.
However, there is a limited speedup of the unstructured sparse model on hardware \cite{park2020optimus,tambe2021edgebert} due to the irregular pattern.
For block-based pruning, the parameter matrix is firstly divided into multiple blocks, and then some unimportant blocks was dropped to reduce storage and computing.
The block-based sparse model having regular pattern can achieve high computational efficiency on hardware.
Nevertheless, the speedup of block-based sparse models \cite{peng2021accelerating} is inefficient since there is limited compression ratio using the block-based pattern.
As for \textit{N:M} group-based structured sparsity, the parameter matrix is divided into multiple groups.
Here we consider consecutive column-wise elements in the matrix gather as a group.
Each group has \textit{M} elements and contains \textit{N} nonzero elements at most.
The \textit{N:M} sparsity can achieve high compression ratio along with computational efficiency on hardware due to its fined-grained regular pattern.
Hence, Transformers with \textit{N:M} sparsity pattern, which remains a lot to be explored, have much more speedup potential than that with unstructured and block-based sparsity.
\begin{table}[hbtp]
\centering
\caption{Comparison between existing three sparsity patterns}
\label{tab:sparsity}
\resizebox{0.45\textwidth}{!}{%
\begin{tabular}{@{}c|ccc@{}}
\toprule
              & Unstructured & Block-based & \textbf{N:M Group-based} \\ \midrule
Visualization &  
    \begin{minipage}[b]{0.3\columnwidth}
		\centering
		\raisebox{-.5\height}{\includegraphics[width=0.6\linewidth]{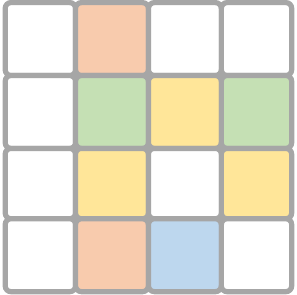}}
	\end{minipage}
	          &
	 \begin{minipage}[b]{0.3\columnwidth}
		\centering
		\raisebox{-.5\height}{\includegraphics[width=0.6\linewidth]{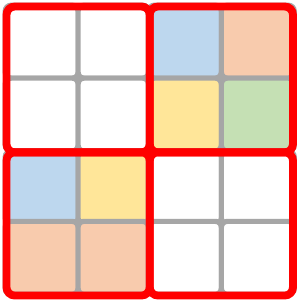}}
	\end{minipage}  
	          & 
    \begin{minipage}[b]{0.3\columnwidth}
		\centering
		\raisebox{-.5\height}{\includegraphics[width=0.6\linewidth]{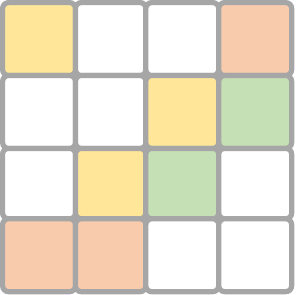}}
	\end{minipage} 
	\\ \midrule
Accuracy      & \textbf{High \checkmark}       & Medium      & \textbf{High \checkmark}          \\ \midrule
Efficiency    & Low        & \textbf{High \checkmark}      & \textbf{High \checkmark}          \\ \midrule
Speedup       & Medium       & \textbf{High \checkmark}      & \textbf{High \checkmark}          \\ \bottomrule
\end{tabular}%
}
\end{table}

\subsection{Pruning Algorithm} \label{subsec:pruning}

Given a pretrained dense Transformer model, generally, a \textit{N:M} sparse Transformer can be trained with the objective as 
\begin{equation}
	\label{equ:def}
	\min_{S(\mathcal{W}, N, M)} \mathcal{L}(\mathcal{W}; \mathcal{D}),
\end{equation}
where $\mathcal{D}$ denotes the observed data, $\mathcal{L}$ represents the loss funtion, $\mathcal{W}$ indicates the parameters of the Transformer, and $S(\mathcal{W}, N, M)$ is the sparse Transformer with \textit{N:M} sparsity pattern. 
$N$ is the number of non-zero values. For dense model $\mathcal{W}$, it can be equivalent to $S(\mathcal{W},  N=M, M)$.

Existing methods NVIDIA ASP\cite{mishra2021accelerating} and SR-STE \cite{zhou2021learning} leverage the single-shot magnitude-based pruning and dynamic sparse training from dense models $\mathcal{W}$ respectively. 
The specific sparse models $S(\mathcal{W}, N, M)$ inherit from global dense models $S(\mathcal{W},  M, M)$ with pre-trained weights and random initialization in ASP and SR-STE. 
This may lead to suboptmial problems, and we observe the ASP and SR-STE hurt the performance significantly on Transformer-based models with the higher sparse ratio (e.g., $\geq$75\%). 
In addition, the ASP and SR-STE undesirably require intensive training computation if we have different hardware constraints with multiple sparsity levels (e.g., \textit{1:8}, \textit{2:8}, \textit{3:8} and \textit{4:8}).

Therefore, we propose a general and simple algorithm for generating models with general \textit{N:M} sparse patterns, namely IDP, which can produce a series of sparse models with different \textit{N:M} configurations. 
Algorithm~\ref{alg:idp} presents the detail of IDP.
To handle the optimization difficulty of the sparse sub-networks inherited from large dense models, we introduce a novel co-training scheme, which optimize different multiple-level \textit{N:M} sparse models simultaneously (e.g., \textit{1:8}, \textit{2:8}, \textit{3:8}, \textit{4:8} and \textit{5:8}). 
During the training phase, we gradually reduce the non-zeros parameters \textit{N}, which can guarantee the super models converge well. We can give the general inheritance mechanism of the IDP as follows:
\begin{equation}
    \small
	\label{equ:inherit}
	S(\mathcal{W}, N_1, M) \leftarrow  S(\mathcal{W}, N_2, M) \leftarrow \cdots \leftarrow S(\mathcal{W}, M, M),
\end{equation}
where $S(\mathcal{W}, N_i, M)$ are \textit{N:M} sparse models, and the $S(\mathcal{W}, M, M)$ represents the dense model, where $ N_1 < N_2 < \cdots < M $, $\leftarrow$ means the smaller model $S(\mathcal{W},  N-1, M)$ prune from $S(\mathcal{W}, N, M)$, named \textbf{inheritance mechanism}.
It requires merely a hyper-parameter $n$ denoted as the end of iterations of $N$.
\RoundOne{
With the novel inheritance mechanism, our IDP training method can be summarized with four steps: 
\begin{itemize}
    \item \underline{Step 1:} Initialize $N=M-1$ and set the dense pretrained model as the first winner model.
    \item \underline{Step 2:} Sparsity inheritance applies the kept parameters of the winner of all the $S(\mathcal{W}, N+1, M)$ candidates to initialize following sparse model $S(\mathcal{W}, N, M)$. 
    \item \underline{Step 3:} Sparse training for \textit{N:M} sparse candidates in several epochs. Parameters are adjusted in every epoch by updating the mask based on the their magnitude. This step generates a new winner model, which is the convergent model at the last epoch.
    \item \underline{Step 4:} If $N=n$, the whole process is finished, or otherwise $N \leftarrow N-1$, and then go to \underline{Step 2}.
\end{itemize}
}

\RoundOne{
We expect to obtain \textit{M-N+1} preserved winner models with different \textit{N:M} sparsity for subsequent deployment.
Additionally, in the forward pass, we leverage the popular group-wise magnitude pruning \cite{mishra2021accelerating,zhou2021learning}.
Parameter matrices are partitioned into multiple groups, every one of which contains $M$ consecutive column-wise elements, as shown in Table~\ref{tab:sparsity}.
And we keep the $N$-largest parameters in these groups and generate corresponding masks $\mathcal{B} \in \{0,1\}^d$.  Specifically, if the $i$-th parameter of $\mathcal{W}$
survived 
in the pruned sub-network, we set $\mathcal{B}_i = 1$, or else $\mathcal{B}_i = 0$.
In the backward pass, recent studies \cite{zhou2021learning,lin2020dynamic} demonstrate that the dynamic sparse training can benefit both model convergence and accuracy, and we follow their methods to calculate gradients. 
}

\begin{algorithm}
	\renewcommand{\algorithmicrequire}{\textbf{Input:}}
	\renewcommand{\algorithmicensure}{\textbf{Output:}}
	\caption{Inherited Dynamic Pruning}
	\label{alg:idp}
	\begin{algorithmic}[1]
	    \REQUIRE Pre-trained dense weights $\mathcal{W}$, datasets $\mathcal{D}$, initial learning rate $\gamma_{0}$ and the end of iterations $n$.
        \FOR{$N=M-1, M-2,.., n$}
           \STATE \RoundOne{\textbf{Sparsity Inheritance}: \\ ~~$S(\mathcal{W}, N, M) \leftarrow$ the winner of $S(\mathcal{W}, N+1, M)$.}
           \FOR{each training iteration $t$}
             \STATE \textbf{Forward Pass}: 
                \\ ~~generate $\mathcal{B}_t$ by group-wise magnitude pruning.
             \STATE \textbf{Backward Pass}: 
                \\ ~~$\mathcal{W}_{t+1} =  \mathcal{W}_t - \gamma_t g( \mathcal{W}_t \odot \mathcal{B}_t) + \lambda((1-\mathcal{B}_t)\odot W_t)$.
           \ENDFOR
        \ENDFOR
	    \ENSURE \RoundOne{A series of \textit{N:M} sparse models with different computation complexity and corresponding masks: the winners of $S(\mathcal{W},  N=M-1, M)$, $S(\mathcal{W},  N=M-2, M)$,.., $S(\mathcal{W},  N=n, M)$.}
	\end{algorithmic} 
\end{algorithm}

\begin{figure} [htbp] 
	\centering
	\includegraphics[width=0.42\textwidth]{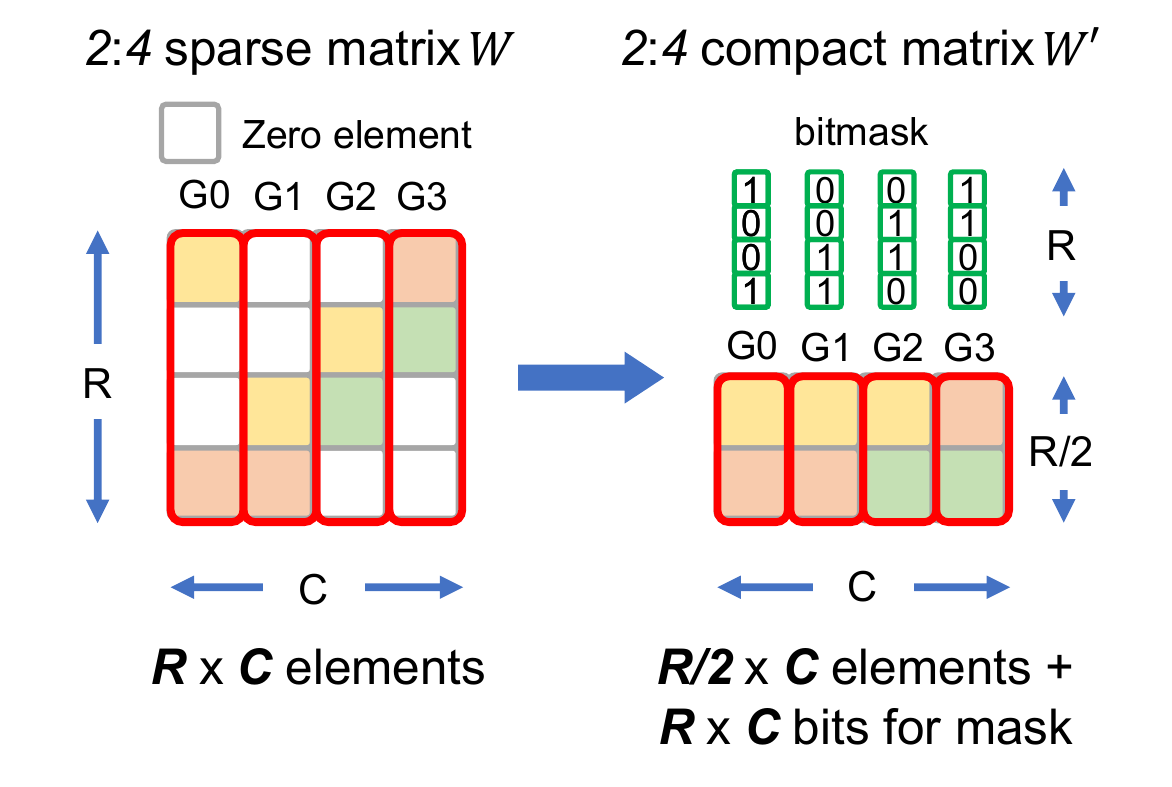}
	\caption{Compact storage scheme example for \textit{N:M} sparse parameter matrix.} 
	\label{fig:compressed_format}
\end{figure}

\subsection{Packing \textit{N:M} Sparse Parameters} \label{subsec:packing}

An \textit{N:M} sparse Transformer can be obtained after IDP, where each group of all parameter matrices only contains at most $N$ non-zero elements.
However, it occupies a large amount of memory since the parameter store scheme is the same as the dense Transformer.
\RoundOne{
We apply the bitmap-based compression scheme to obtain a compact \textit{N:M} sparse Transformer, which greatly achieves saving on storage for deployment.
Compared to COO, CSC, CSR \cite{hoefler2021sparsity}, and step indexing \cite{zhang2016cambricon}, our scheme has better compression performance in the range of practical \textit{N:M} sparsity.}
Fig.~\ref{fig:compressed_format} presents the compression scheme of \textit{N:M} sparse parameter matrix using \textit{2:4} sparsity as an example.
For a parameter matrix $W\in\mathbb{R}^{R\times C}$, after IDP, there are at most \textit{2} non-zero elements in a group.
The entire parameter matrix has $\frac{R}{2} \times C$ remaining non-zero elements.
In our scheme, we merely preserve non-zero elements in each group, and use a binary mask to indicate the elements' position.
By using our scheme, a \textit{2:4} parameter matrix $W\in\mathbb{R}^{R\times C}$ can be stored with $\frac{R}{2} \times C$ valid elements and $R \times C$ bits for the mask, instead of the $R \times C$ elements.

Considering a dense parameter matrix $W\in\mathbb{R}^{R\times C}$, in which all elements are quantized using $q$ bits.
$W$ can be compressed to $\tilde{W}\in\mathbb{R}^{R \times \lceil \frac{C}{M} \rceil N}$, where there are $R\lceil \frac{C}{M} \rceil$ groups and each group has $N$ non-zero parameters at most.
The storage requirement of $W$ is $qRC$ bits, and after pruning, we can only store the \textit{N:M} sparse matrix $\tilde{W}$ in a compact way with only  $qR\lceil\frac{C}{M}\rceil N$ bits, and an additional binary mask with $RC$ bits.
Therefore, the compression ratio (CR) can be represented as: 

\begin{equation}
	CR = \frac{qC}{q\lceil\frac{C}{M}\rceil N+C}.
	\label{equ:compression_ratio}
\end{equation}

\section{Hardware Optimization} \label{sec:hwarch}


The flexible and efficient hardware architecture, namely STA, is developed for \textit{N:M} sparse Transformers in this section.
We first present the overall architecture of STA, and then elaborate on designs of its core computing engines, including DMME and scalable softmax module.

\begin{figure} [hbtp] 
	\centering
	\includegraphics[width=0.46\textwidth]{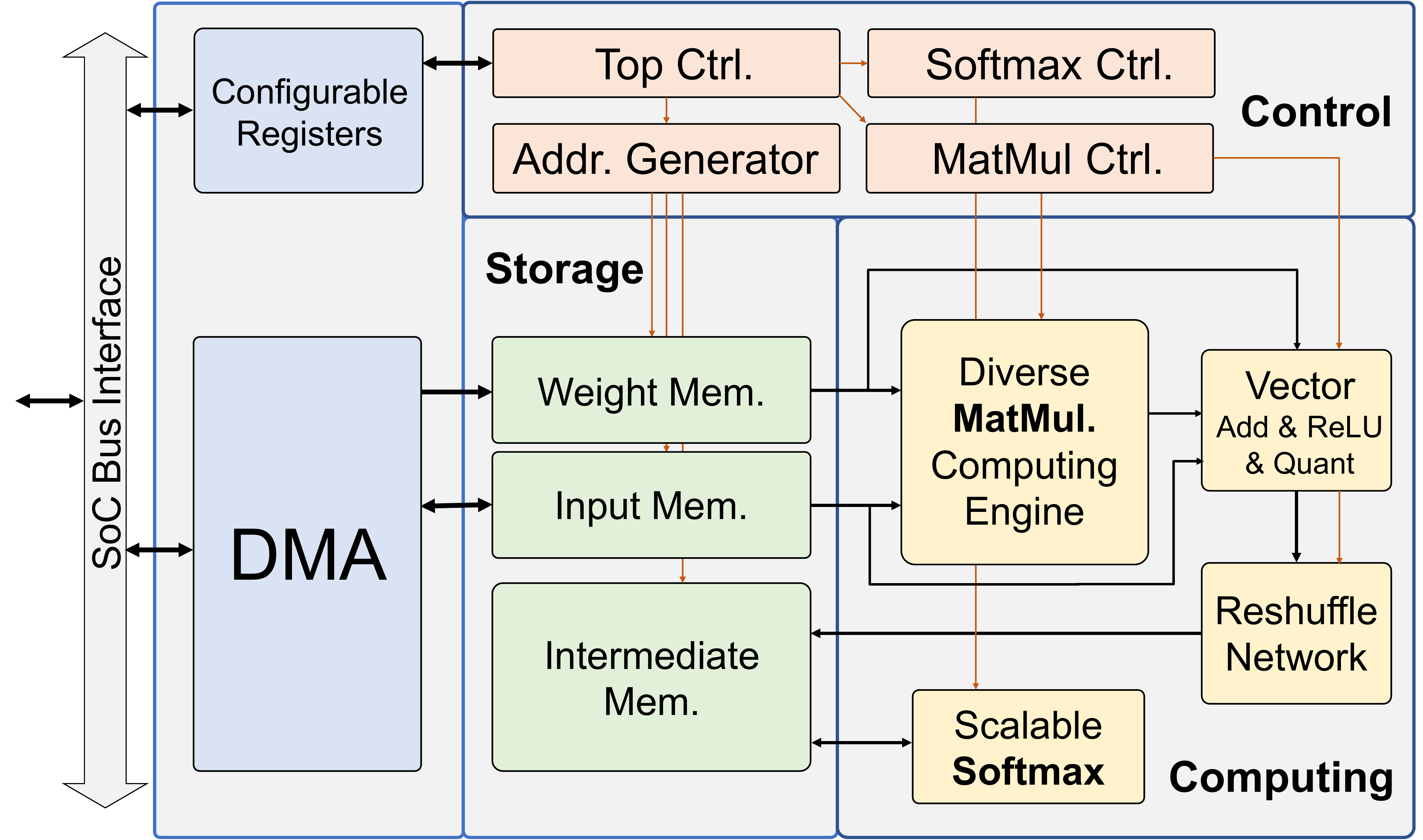}
	\caption{The overall architecture of STA. It is composed of computing, storage, and control function blocks. These red arrows pass control signals, while those black arrows transfer data.} 
	\label{fig:arch}
\end{figure}

\subsection{Overall Architecture} \label{subsec:ova}
The overall architecture of STA is shown in Fig.~\ref{fig:arch}, which consists of three major function blocks, including computing, storage, and control. 
The computing blocks consist of a diverse MatMul computing engine , namely DMME, a scalable softmax module, a vector unit, and a data reshuffle network.
Dominated operations of \textit{N:M} sparse Transformers, i.e. sparse-dense or dense-dense MatMuls, are performed by DMME on-the-fly with the dynamic configuration under high computational efficiency.
The scalable softmax module is responsible for the softmax operation in MHA ResBlocks, eliminating the off-chip transfer for intermediate data.
The vector unit takes charge of operations with low computational density including bias addition, residual addition, and activation functions.
The reshuffle network reorders the temporary results before writing back to the intermediate on-chip memory.
As for on-chip storage, it can be partitioned into three parts, including the weight memory, the input memory, and the intermediate memory. 
The weight and input memory store model parameters and input data of Transformers from the off-chip memory, respectively.
The results of a ResBlock are also written back to the input memory, and pass to the off-chip memory.
And all the temporary results in a ResBlock will be stored in the intermediate memory with no communication to the external memory.

\begin{figure} [hbtp] 
	\centering
	\includegraphics[width=0.48\textwidth]{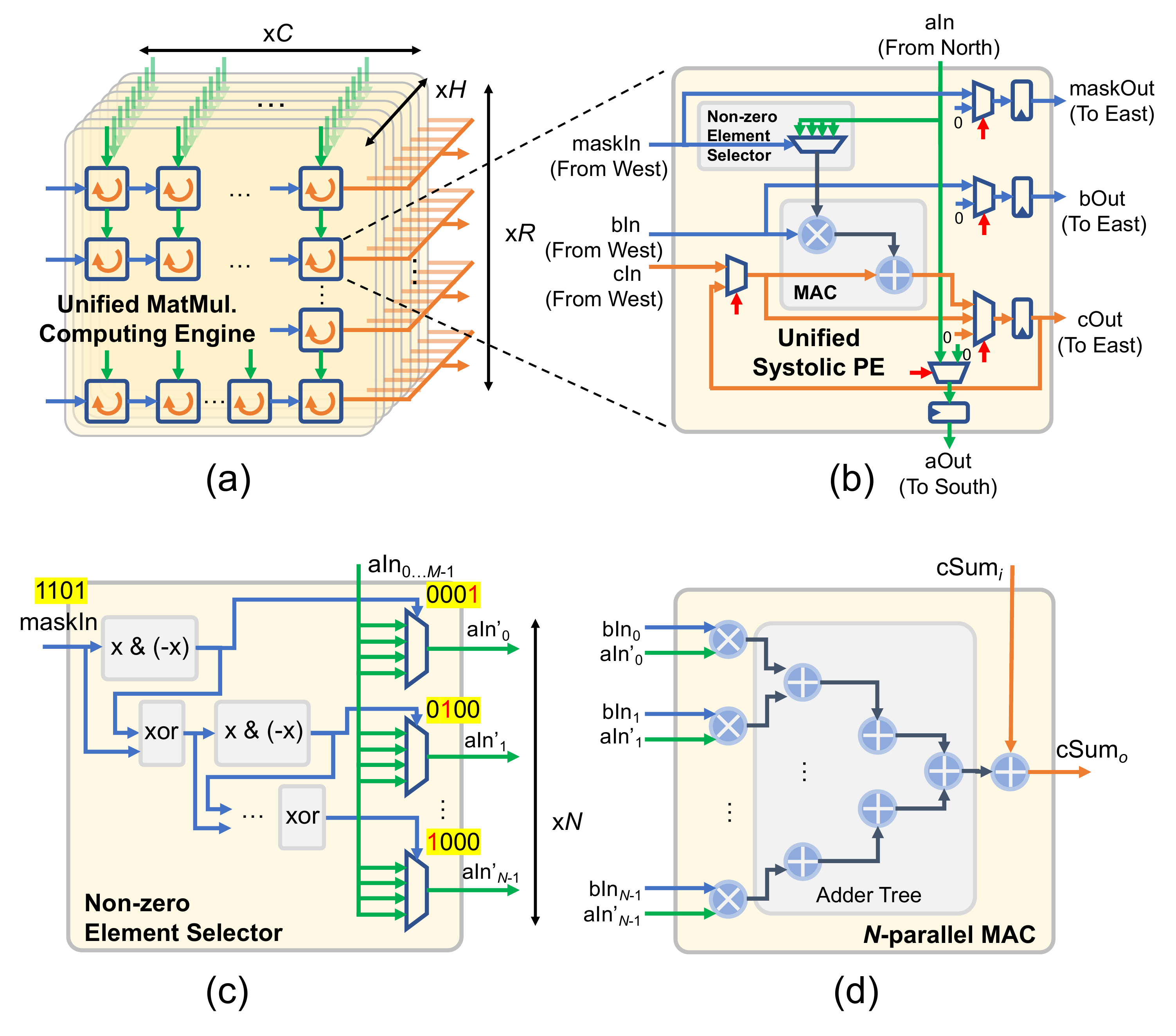}
	\caption{The hierarchical architecture of DMME. It consists of $H$ parallel unified MatMul computing engine (a). Each engine contains $R \times C$ unified systolic PE capable of handling both sparse-dense and dense-dense dot products (b). The key components of PEs, non-zero element selector and the \textit{N}-parallel MAC, are in (c) and (d), respectively.} 
	\label{fig:mircoarch}
\end{figure}

\subsection{DMME} \label{subsec:DMME}
DMME unifies both sparse-dense and dense-dense MatMuls with a high computational efficiency in \textit{N:M} sparse Transformers. 
When it performs sparse-dense MatMuls, it merely loads nonzero weight parameters and selects corresponding activations to compute, thereby improving computational efficiency. 

The architecture of the DMME is illustrated in Fig.~\ref{fig:mircoarch}. 
It is a two-level hierarchy design with a full exploration of parallelism inside the MatMuls of \textit{N:M} sparse Transformers.
The exploited parallelism consists of head parallelism, row parallelism, and column parallelism, which are denoted as $H$, $R$, and $C$, respesctively.
The DMME is composed of $H$ parallel $R\times C$ unified MatMul computing engine (Fig.~\ref{fig:mircoarch}~(a)), every one of which can efficiently realize both sparse-dense and dense-dense MatMuls in a time-division multiplexing manner.
The capability of performing sparse-dense and dense-dense MatMuls comes from the inner unified systolic PE (Fig.~\ref{fig:mircoarch}~(b)).
It is composed of a non-zero element selector, a \textit{N}-parallel MAC, multiple multiplexers, and registers.
The non-zero element selector, only being activated in the sparse-dense MatMul mode, is to select the proper activation according to the input bitmask.
The \textit{N}-parallel MAC accepts \textit{N} \textit{16}-bit input data and parameters, realizes inner product, and then accumulates the result with the local \textit{32}-bit output partial sum.
The multiplexers and registers are used for datapath selection and temporary data storage, respectively.
The design of non-zero element selector is presented in Fig.~\ref{fig:mircoarch}~(c).
It takes as input an \textit{M}-bit mask, in which only \textit{N} bits are set as \textit{1} to indicate the position of non-zero elements, and then generates \textit{N} one-hot encoding masks to select the corresponding \textit{N} data for dot product computation.
The translation to \textit{N} one-hot encoding masks is performed by cascading the simple bit-arithmetic blocks and \textit{XOR} gates.
With the help of \textit{N} one-hot encoding masks, data related to non-zero parameters are fed into the \textit{N}-parallel MAC along with these non-zero parameters in one group.
\RoundOne{It could be pointed out that the non-zero element selector can be further optimized by pruning the redundant indexing indicators and element candidates.}
The \textit{N}-parallel MAC, as shown in Fig.~\ref{fig:mircoarch}~(d), is composed of \textit{N} parallel multipliers, an adder tree, and a final accumulator.
Both non-zero element selector and \textit{N}-parallel MAC are fully pipelined to maximize the throughput of DMME.


\begin{figure} [htbp] 
	\centering
	\includegraphics[width=0.48\textwidth]{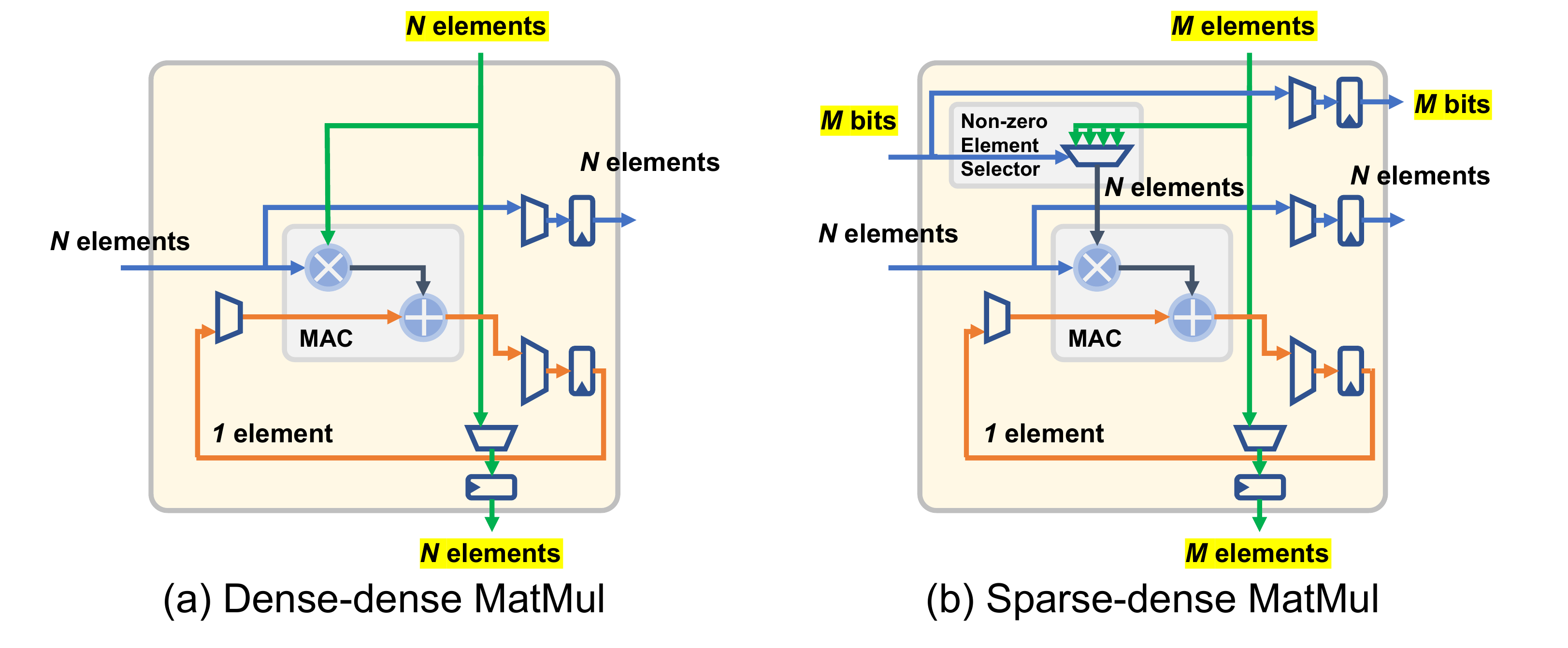}
	\caption{The activated datapath of PEs under (a) dense-dense and (b) sparse-dense modes.} 
	\label{fig:datapath}
\end{figure}

The activated datapath under dense-dense and sparse-dense MatMuls are presented in Fig.~\ref{fig:datapath}~(a) and (b), respectively.
The dense-dense mode of PEs would be only activated when performing the self-attention operation of MHA.
Both input operands are arranged in dense sequences in this mode.
In this case, \textit{N} elements, as an operand, in the input sequences are in parallel streamed into the PE from the west and the north, respectively.
In a cycle, the PE performs a dot product with a size of \textit{N} under dense-dense mode.
The partial sum is stored in the local registers, and the input operands from the west and the north stored in the registers are passed into adjacent PEs on the east and south, respectively.
For energy saving, the non-zero element selector is bypassed to avoid signal switching.
Under sparse-dense MatMul mode, as shown in Fig.~\ref{fig:datapath}~(b), the input operands is different from that under the dense-dense mode.
In a cycle, \textit{N} non-zero parameters in a group along with the corresponding $M$-bit mask are streamed into the PE from the west, while $M$ data in one group are fed from the north.
The \textit{N} valid data in pair with the input parameters is picked up by the non-zero element selector, and then performs a dot product with these input parameters.
When the computing task is done, under either dense-dense or sparse-dense modes, the PE turns into the shifting mode, accepts the results from its western PE to its local registers and transfers its local result registers to the east.

\begin{figure} [hbtp] 
	\centering
	\includegraphics[width=0.48\textwidth]{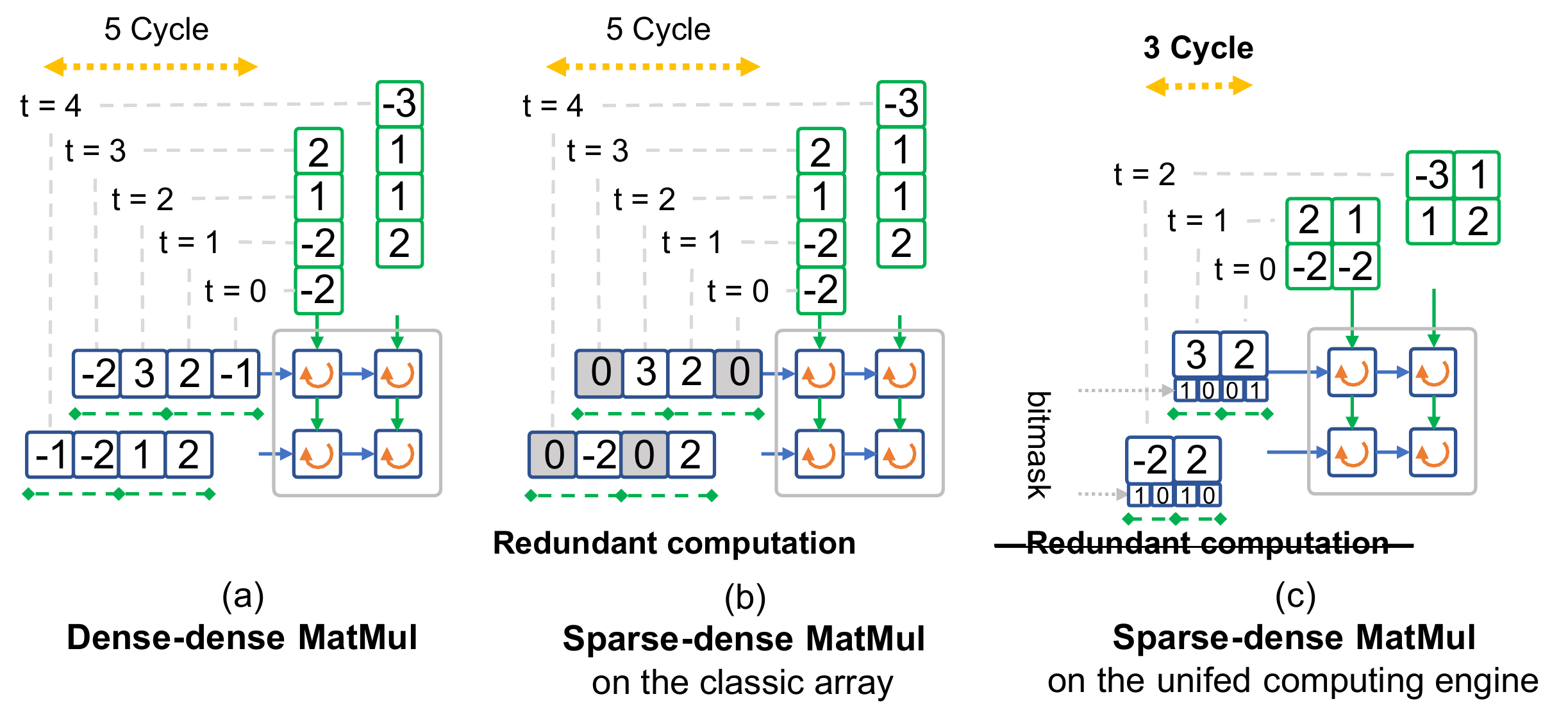}
	\caption{\RoundOne{Computing dataflows of DMME when it performs (a) dense-dense and (c) sparse-dense MatMuls. Compared to (b) as a baseline, (c) eliminates all zero-valued redundant operations under sparse-dense mode, thus improving computational efficiency.}}
	\label{fig:dmme_dataflow}
\end{figure}

\subsection{\RoundOne{Supporting Efficient Matrix Computations}}

\RoundOne{
STA is capable of supporting both sparse-dense and dense-dense MatMuls of \textit{N:M} sparse Transformers in an efficient way.
We demonstrate this significant capability of STA by exploiting four aspects: the computing dataflow of DMME, data access pattern of DMME, data mapping of input memory, and datapath from input memory to DMME.
}

\RoundOne{
Fig.~\ref{fig:dmme_dataflow} illustrates efficient computing dataflows of DMME under both dense-dense and sparse-dense modes.
For simplicity, we assume \textit{N:M} is \textit{1:2}, and MatMul is performed by $2 \times 4$ input sequences and $4 \times 2$ parameter sequences.
Here, we consider the computing engine in \cite{lu2020hardware} as a baseline, which is orchestrated as a classic systolic array.
In Fig.~\ref{fig:dmme_dataflow}~(a), DMME finishes the dense-dense MatMul in the given computing task using \textit{5} cycles, which consumes the same cycles as the baseline.
Hence, DMME achieves the same computational efficiency as the baseline when performing dense-dense MatMuls.
As for sparse-dense MatMuls, Fig.~\ref{fig:dmme_dataflow}~(b) and (c) present the computational manner of the baseline and DMME, respectively.
The baseline takes \textit{5} cycles to finish the task, while it cost merely \textit{3} cycles by DMME since the redundant operations can be skipped with no waste on computing cycles.
For sparse-dense MatMuls, DMME improves the computational efficiency by eliminating redundant computations, thereby significantly reducing latency and energy.
}

\begin{figure} [htbp] 
	\centering
	\includegraphics[width=0.48\textwidth]{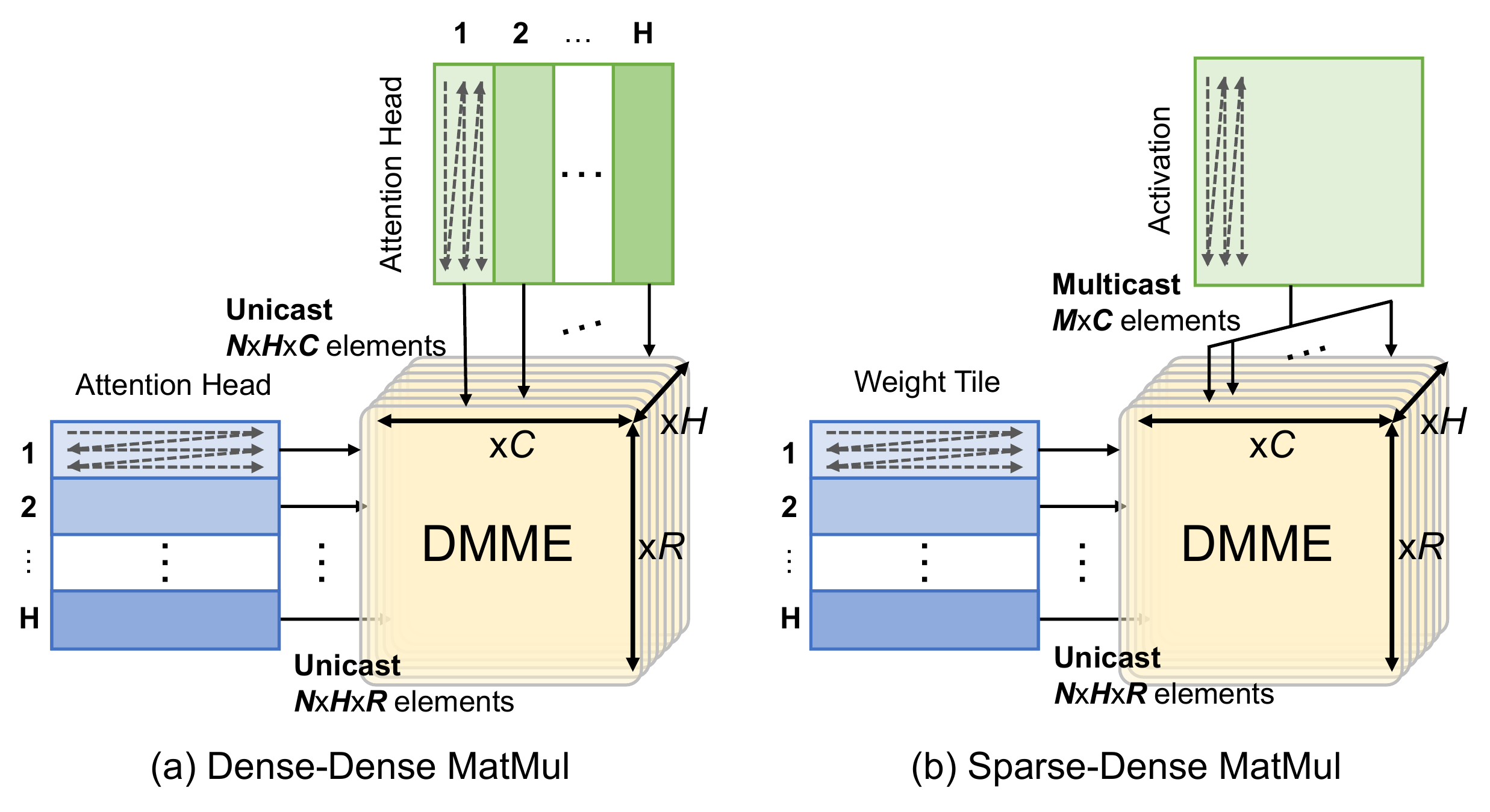}
	\caption{Data access pattern of DMME to support efficient MatMuls under (a) dense-dense and (b) sparse-dense modes.}
	\label{fig:dmme_pattern}
\end{figure}

\RoundOne{
Fig.~\ref{fig:dmme_pattern}~(a) and (b) presents data access patterns of DMME when it
performs dense-dense and sparse-dense MatMuls, respectively.
For dense-dense MatMuls, attention heads as input are both separated into $H$ tiles.
In this case, DMME can be decomposed as $H$ independent systolic arrays, every one
of which fetches elements from the corresponding tiles to the top-most and left-most PEs, respectively.
For sparse-dense MatMuls, compressed weight parameters are divided into $H$ tiles.
Every cycle DMME fetches $NR$ weight elements from all $H$ tiles in parallel, and casts them one-on-one to the left-most systolic PEs in $H$ systolic arrays. 
DMME is also required to access $MC$ activation elements, and broadcasts them to the top-most PEs in all $H$ unified systolic arrays.
}

\begin{figure} [htbp] 
	\centering
	\includegraphics[width=0.48\textwidth]{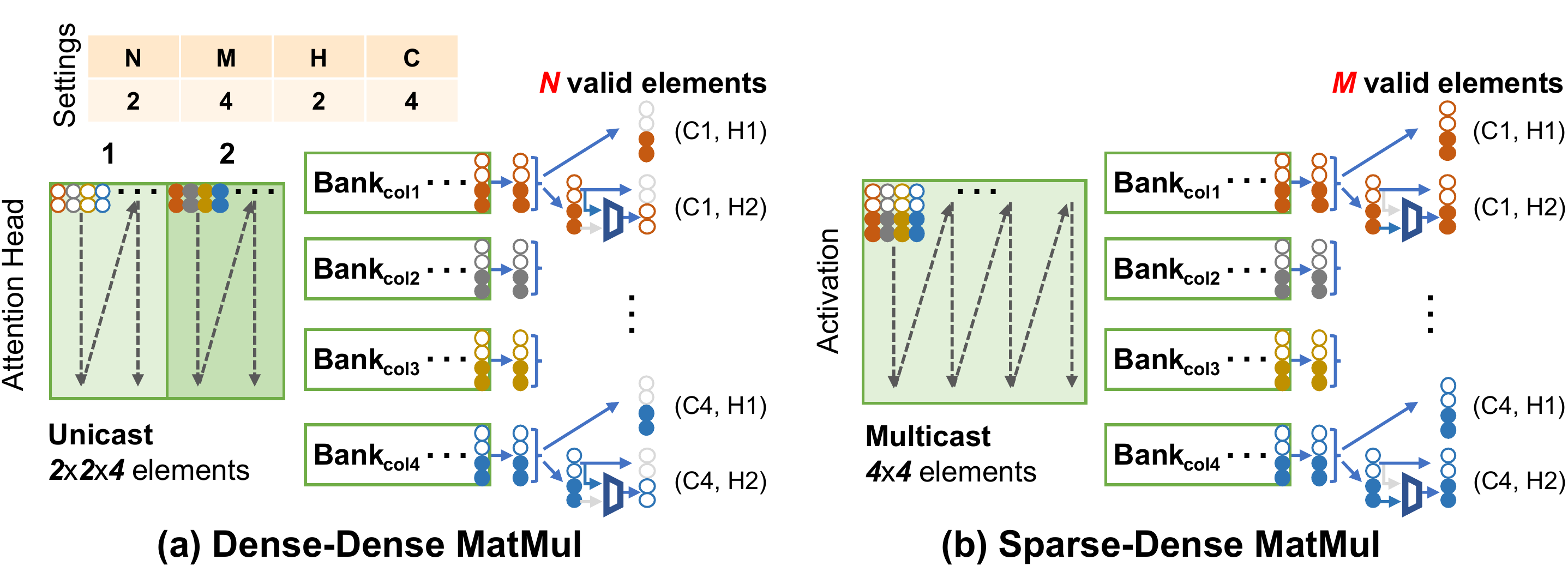}
	\caption{\RoundOne{Data mapping of input memory and datapath from input memory to DMME under (a) dense-dense and (b) sparse-dense modes.}}
	\label{fig:input_mem}
\end{figure}

\RoundOne{
To balance the bandwidth of input memory when switching between dense-dense and sparse-dense modes, we make $NH$ equal to $M$ of STA.
There are $C$ banks of STA for input data storage.
Fig.~\ref{fig:input_mem} illustrates data mapping of input memory and datapath from input memory to DMME by assuming $N$ is 2, $M$ is 4, $H$ is 2, and $C$ is 4.
The data storage storage structure in the input memory is varied for different computing modes.
In dense-dense mode, DMME performs 2 parallel dense-dense MatMuls for attention mechanism of Transformers. 
There are 2 tiles for the loaded input data. 
As shown in Fig.~\ref{fig:input_mem}~(a), the first bank is connected to the first column of DMME, and the first address of the bank indexes the data from the first 2 elements at the first column in the tile one and two, respectively. 
In sparse-dense mode, DMME performs MatMuls with the \textit{N:M} sparse parameters. 
As depicted in Fig.~\ref{fig:input_mem}~(b), we do not tile input data for sparse-dense MatMul.
The first address of bank one that connected to the first column of
DMME, indexes the data from the first 4 elements at the first column of input data.
It is the same of the indexing principle for the other banks in both computing modes.
}

\RoundOne{
As for the datapath from input memory to DMME, we takes the first column of DMME as an example.
There are 4 elements accessed from the first bank streaming to the first column with a head dimension of 2.
In dense-dense mode, as shown in Fig.~\ref{fig:input_mem}~(a), the unified systolic PE in the first head of the first column directly receives these 4 elements, and the lowest 2 elements are fed into \textit{N}-parallel MAC for computing.
However, the unified systolic PE in the second head of the first column requires 2-to-1 MUXs to select the correct 2 elements from the 4 accessed elements for computing.
In sparse-dense mode, as presented in Fig.~\ref{fig:input_mem}~(b), data accessed from the first bank broadcast to unified systolic PEs in all head dimensions of the first column of DMME.
}

\subsection{Scalable Softmax Module} \label{subsec:softmax}

\begin{figure} [hbtp] 
	\centering
	\includegraphics[width=0.48\textwidth]{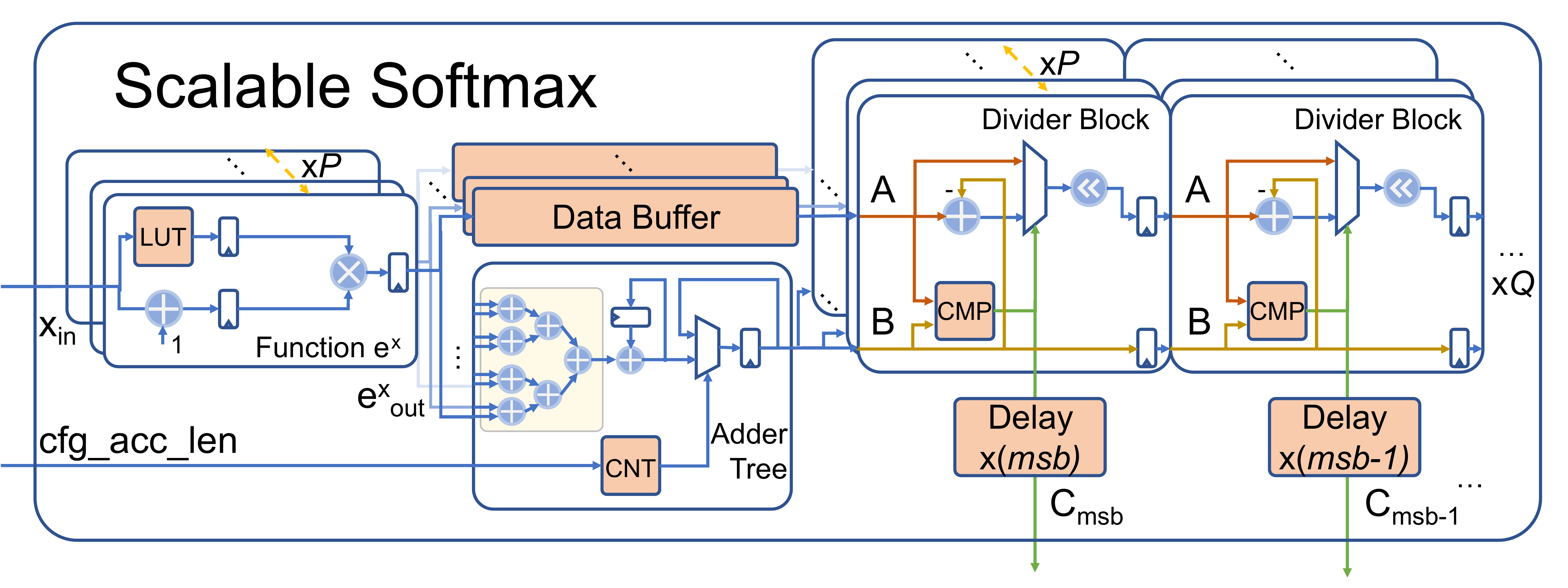}
	\caption{The architecture of the scalable softmax operator.} 
	\label{fig:softmax}
\end{figure}

The softmax function takes as input a vector $\textbf{x}$ of $n$ real numbers, and normalizes it into a probability distribution consisting of $n$ probabilities proportional to the exponentials of the input numbers.
It is critical for Transformer dedicated accelerators to contain a softmax hardware implementation since the softmax function appears in every MHA module of Transformers.
Fig.~\ref{fig:softmax} presents the details of our proposed scalable softmax architecture, which is capable of performing softmax functions of arbitrary length. 
It keeps all the intermediate results fully local, avoiding off-chip data communication.

The architecture has two adjustable parameters, $P$ and $Q$, where $P$ denotes the parallelism of the architecture, and $Q$ represents the pipeline depth, as well as the output precision.
$P$ input data are streamed into the softmax module in parallel, and transformed into the exponent outputs.
The exponent outputs are not only temporarily stored in the data buffer, but also used as input for further accumulation.
Once the accumulation process is done, the divider module takes both accumulated results and exponent outputs as input to perform $Q$-level pipelined division, and generates $P$ softmax function outputs represented by $Q$-bit.

As shown in Fig.~\ref{fig:softmax}, the scalable softmax module consists of three major parts: an area-efficient exponential function, a partial sum accumulator, and a scalable divider.
The exponential function is approximated using a lookup table combined with a first-order Taylor expansion.
An exponential operator can be implemented using only one multiplier and one adder.
The configurable partial sum accumulator can adapt to input vectors of various lengths, which improves the flexibility of the hardware.
To reduce the latency of the division, we design a highly parallel divider by cascading multiple divider blocks with pipelines, where a divider block is composed of subtractors and shifters with little cost on hardware.

\section{Experimental Results} \label{sec:res}




In this section, we comprehensively evaluate both algorithm and hardware optimizations of the proposed framework.
Three benchmark sets with varying size and complexity are applied to evaluate the proposed framework.

\subsection{Experimental Setup}

\subsubsection{\textbf{Benchmark Sets}}
~
\newline
\indent The first set focuses on the evaluation of algorithm optimizations, by comprehensively presenting improvements on both model accuracy and compression ratio under various \textit{N:M} configurations.
This benchmark set comprises a BERT model \cite{kenton2019bert}, the well-known Transformer-based model, and four evaluation datasets from the GLUE benchmark \cite{wang2018glue}, including WNLI, QNLI, QQP, and MRPC.
WNLI is a reading comprehension task.
QNLI is a question-answering dataset consisting of question-paragraph pairs.
QQP is a collection of question pairs from the community question-answering website Quora. 
MPRC is a corpus of sentence pairs automatically extracted from online news sources.
For these four tasks, we report accuracy of the validation sets.
We also report the compression ratio on BERT by setting various \textit{N:M} configurations and quantized bitwidth of parameters.

The second set dives into hardware resource consumption. 
Firstly, we study consumption of DMMEs, the core computing engine in the STA, at any common scale of \textit{N:M} sparsity, and then we explore hardware utilization of representative STAs under various FPGA devices.
For the former evaluation, DMMEs are not allowed to be synthesized using DSP blocks, which can be done by setting the property \textit{MAX\_DSP} as zero.
Hence, the consumption of LUTs and FFs can measure the cost of combinational logic and sequential logic for DMMEs, respectively.
The utilization of LUTs and FFs are reported at the synthesis stage as the metric of hardware resource requirements.
For the latter evaluation, the resource utilization of STAs on various FPGA devices are reported at the implementation stage, including the consuming amount of LUTs, FFs, BRAMs, and DSPs.

The third set studies performance improvements of overall STA hardware system on multiple FPGA platforms when deploying various Transformer-based models.
All key configurations of evaluated models in benchmark sets are presented in Table~\ref{tab:benchmark}.
At first, we evaluate the processing time with a single batch on all MHA and FFN ResBlocks in varying models from TinyBERT \cite{jiao2020tinybert}, Dino \cite{caron2021emerging}, and the Transformer-base model \cite{vaswani2017attention}.
The selected models target different applications.
TinyBERT is a lightweight BERT model for many language tasks.
Dino, a tiny vision Transformer, can be the backbone for a lot of computer vision tasks.
Transformer-base model is the classic one for the neural machine translation (NMT) task.
Considering NMT is one of the sequence-to-sequence tasks, hence we split the Transformer-base model into two parts, the stacked encoders and decoders, respectively.
We finally make a fair comparison of the implemented STAs with previous works and commercial products using a shallow Transformer, which is the commonly used benchmark model in \cite{li2020ftrans,peng2021accelerating}.
Latency, throughput, power, energy efficiency and MAC efficiency are key metrics for applications, and thus used for performance evaluation.


\begin{table}[htbp]
\centering
\caption{Key configurations of Transformer-based models in benchmark sets}
\label{tab:benchmark}
\resizebox{0.48\textwidth}{!}{%
\begin{tabular}{@{}cccccccc@{}}
\toprule
Benchmark                & Model                                                                       & \begin{tabular}[c]{@{}c@{}}Num. of \\ Encoders\end{tabular} & \begin{tabular}[c]{@{}c@{}}Num. of\\ Decoders\end{tabular} & \begin{tabular}[c]{@{}c@{}}Sequence \\ length\end{tabular} & \begin{tabular}[c]{@{}c@{}}Attention \\ heads\end{tabular} & \begin{tabular}[c]{@{}c@{}}Hidden\\ size\end{tabular} & \begin{tabular}[c]{@{}c@{}}Intermediate\\ size\end{tabular} \\ \midrule
Set I                    & BERT                                                                        & 12                                                          & 0                                                          & 128                                                              & 12                                                         & 768                                                   & 3072                                                        \\ \midrule
\multirow{10}{*}{Set III} & TinyBERT4                                                                   & 4                                                           & 0                                                          & 128                                                              & 12                                                         & 312                                                   & 1200                                                        \\ \cmidrule(l){2-8} 
                         & Dino-vits8                                                                  & 12                                                          & 0                                                          & 64                                                               & 6                                                          & 384                                                   & 1536                                                        \\ \cmidrule(l){2-8} 
                         & \begin{tabular}[c]{@{}c@{}}Transformer-base\\ stacked encoders\end{tabular} & 6                                                           & 0                                                          & 64                                                               & 8                                                          & 512                                                   & 2048                                                        \\ \cmidrule(l){2-8} 
                         & \begin{tabular}[c]{@{}c@{}}Transformer-base\\ stacked decoders\end{tabular} & 0                                                           & 6                                                          & 64                                                               & 8                                                          & 512                                                   & 2048                                                        \\ \cmidrule(l){2-8} 
                         & \begin{tabular}[c]{@{}c@{}}Shallow\\ Transformer\end{tabular}               & 2                                                           & 1                                                          & 64                                                               & 4                                                          & 200                                                   & 800                                                         \\ \bottomrule
\end{tabular}%
}
\end{table}

\subsubsection{\textbf{Implementation Details}} \label{subsec:impl}
~
\newline
\indent For algorithm implementation (Set I), the pre-trained models, the scripts and datasets are provided by the HuggingFace repository \cite{wolf2019huggingface}.
All models are implemented and executed using PyTorch v1.5.

As for hardware implementation (Set II \& III), all modules of STA are designed in synthesizable SystemVerilog with the aid of hardware components from the BaseJump standard template library \cite{taylor2018basejump} and the PULP platform \cite{rossi2015pulp}.
Xilinx Vivado 2018.2 is the tool for synthesis and implementation.
We implement STA on three types of FPGA devices with various scales, including Xilinx ZYNQ Z7020 (XC7Z020), Xilinx Virtex-7 FPGA (XC7VX485T), and Xilinx UltraScale+ FPGA (XCVU13P).
Specifically, XC7Z020 is low-cost and low-resource System-on-Chip device equipped with a dual-core ARM Cortex-A9 processor and FPGA, which is fabricated in the 28 nm technology node.
XC7VX485T is a relatively large FPGA device fabricated in the 28 nm technology node.
XCVU13P, fabricated in the 16 nm technology node, is an extremely expensive and advanced FPGA device with abundant hardware resource.

\begin{figure*}[htb]
    \centering
    \subfloat[MNLI]{\includegraphics[width=0.21\linewidth]{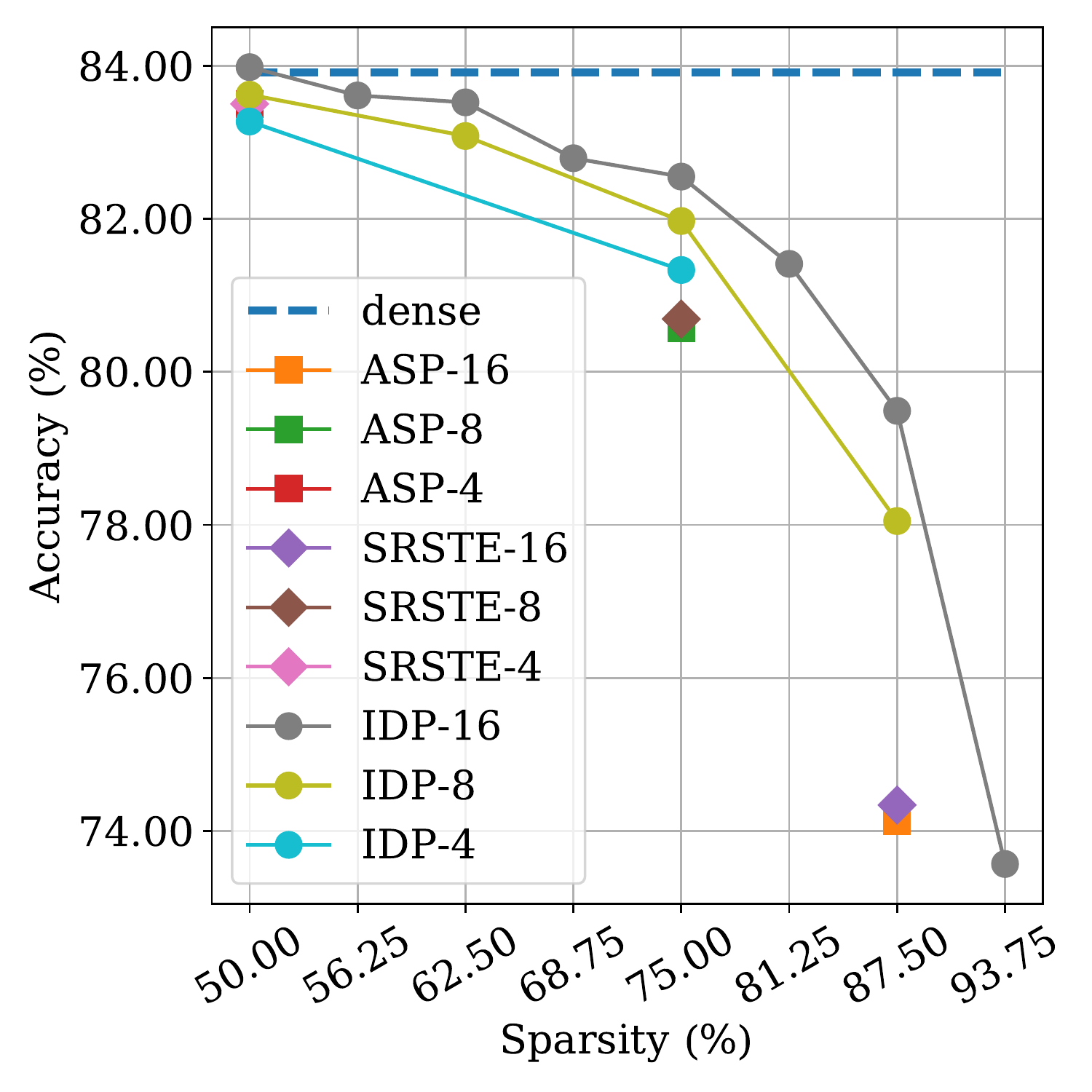}}
    \hfill
    \subfloat[QNLI]{\includegraphics[width=0.21\linewidth]{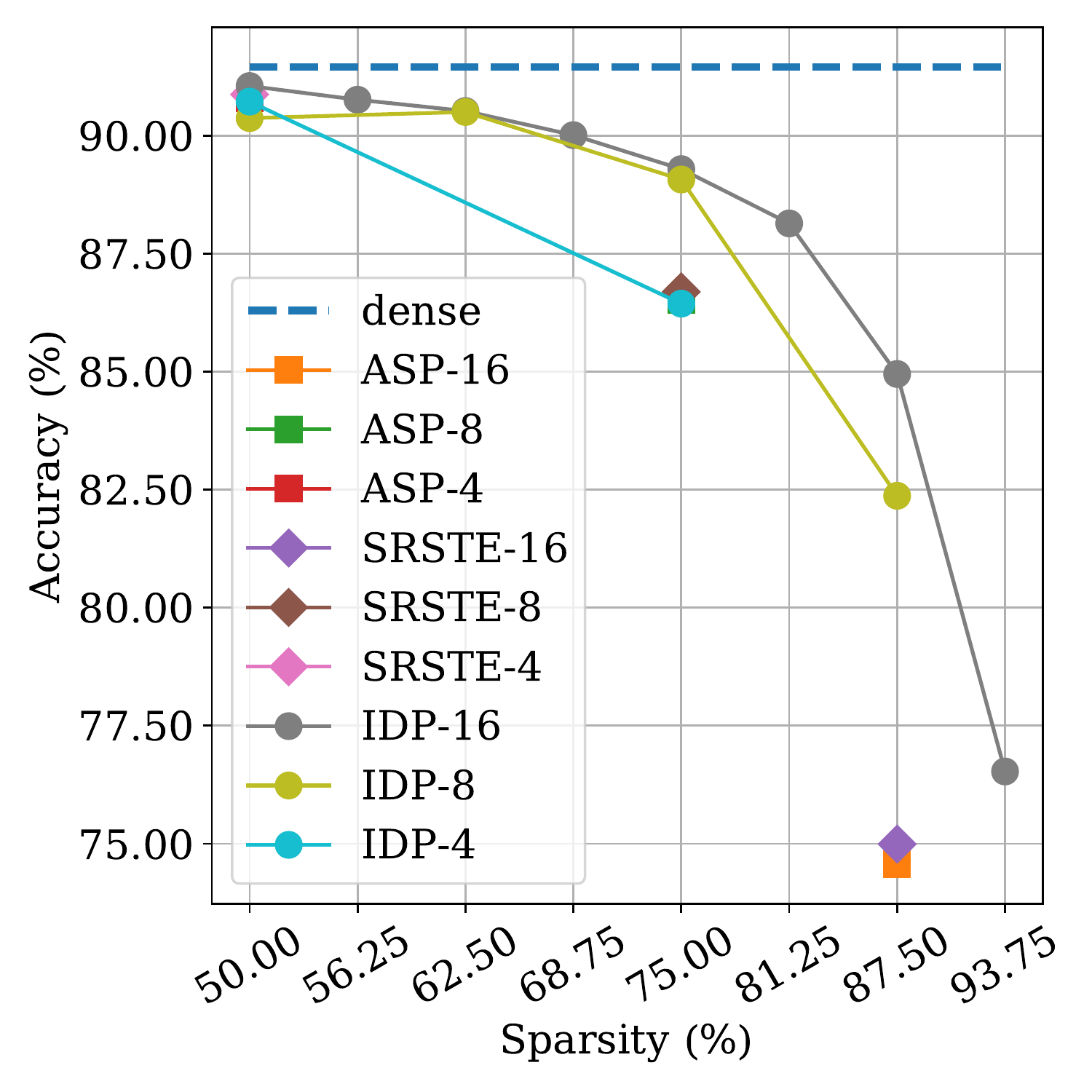}}
    \hfill
    \subfloat[QQP]{\includegraphics[width=0.21\linewidth]{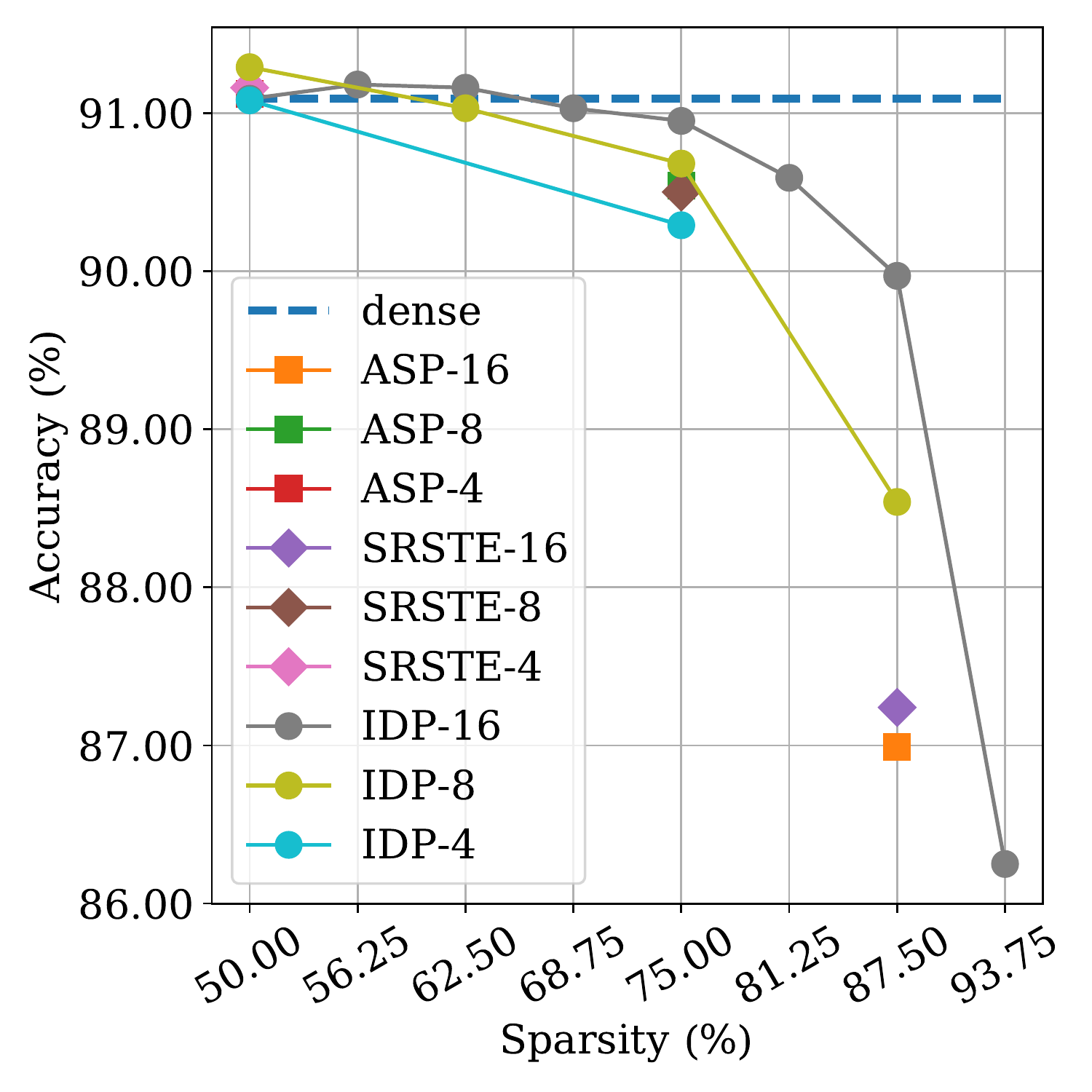}}
    \hfill
    \subfloat[MRPC]{\includegraphics[width=0.21\linewidth]{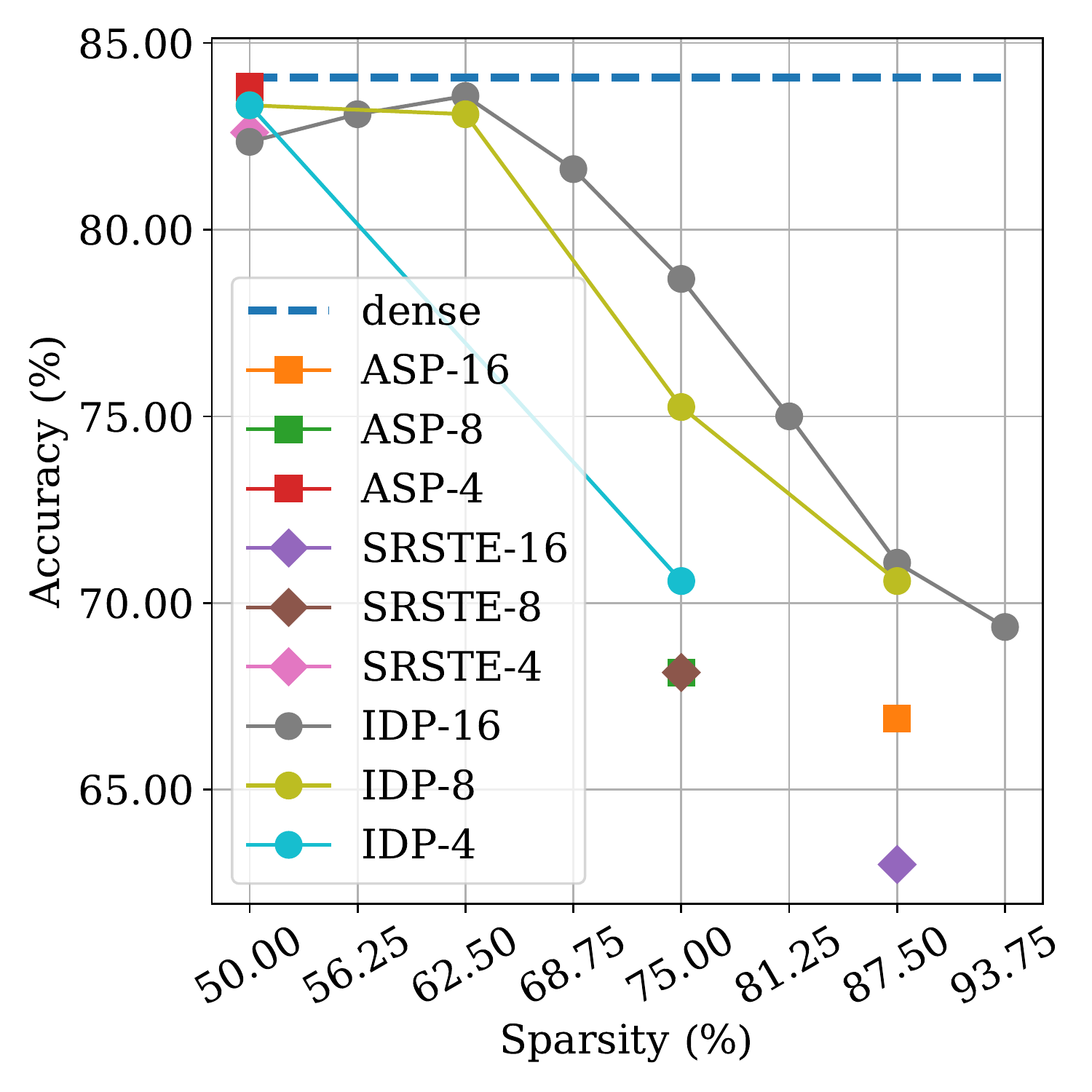}}
    \caption{Pruning results on various tasks incluing (a) MNLI, (b) QNLI, (c) QQP, and (d) MRPC in comparison with ASP \cite{mishra2021accelerating} and SR-STE \cite{zhou2021learning}.}
    \label{fig:res_acc}
\end{figure*}

\subsection{Benchmark Set I: Algorithm Optimizations}

For benchmark set I, ASP \cite{mishra2021accelerating} and SR-STE \cite{zhou2021learning}, the two existing methods for acquiring \textit{N:M} sparse models, are selected as our baselines. 
The reported accuracy of baselines is obtained by training with released open-source code.
For a fair comparison, the generated \textit{N:M} sparse models using ASP, SR-STE, and our method are achieved with identical finetune epochs.
For all tasks, we use a batch size of 32 and a initial learning rate of 2e-5.
For WNLI, QNLI, QQP, there are 3 epochs to recover accuracy for every step of N, while there are 5 epochs for MRPC.


Compared with existing methods, Fig.~\ref{fig:res_acc} shows that IDP can achieve comparable or better accuracy under 75.00\% sparse ratio. In addition, the IDP can outperform the ASP and SR-STE method significantly with the sparse ratio increases. For instance, we can find that under 87.50\% (\textit{2:16}) sparse ratio, IDP consistently obtains large performance improvements to the baseline on all tasks ($5.36\%$ accuracy gain on MNLI, $10.38\%$ accuracy gain on QNLI, $2.98\%$ accuracy gain on QQP, and $8.08\%$ accuracy gain on MRPC). Therefore, we can obtain the state-of-the-art \textit{N:M} sparse models for FPGA-based platform deployment with the plug-and-play IDP algorithm. 

\begin{figure}[htb]
    \centering
    \subfloat[Bitwidth = 4bit]{\includegraphics[width=0.23\textwidth]{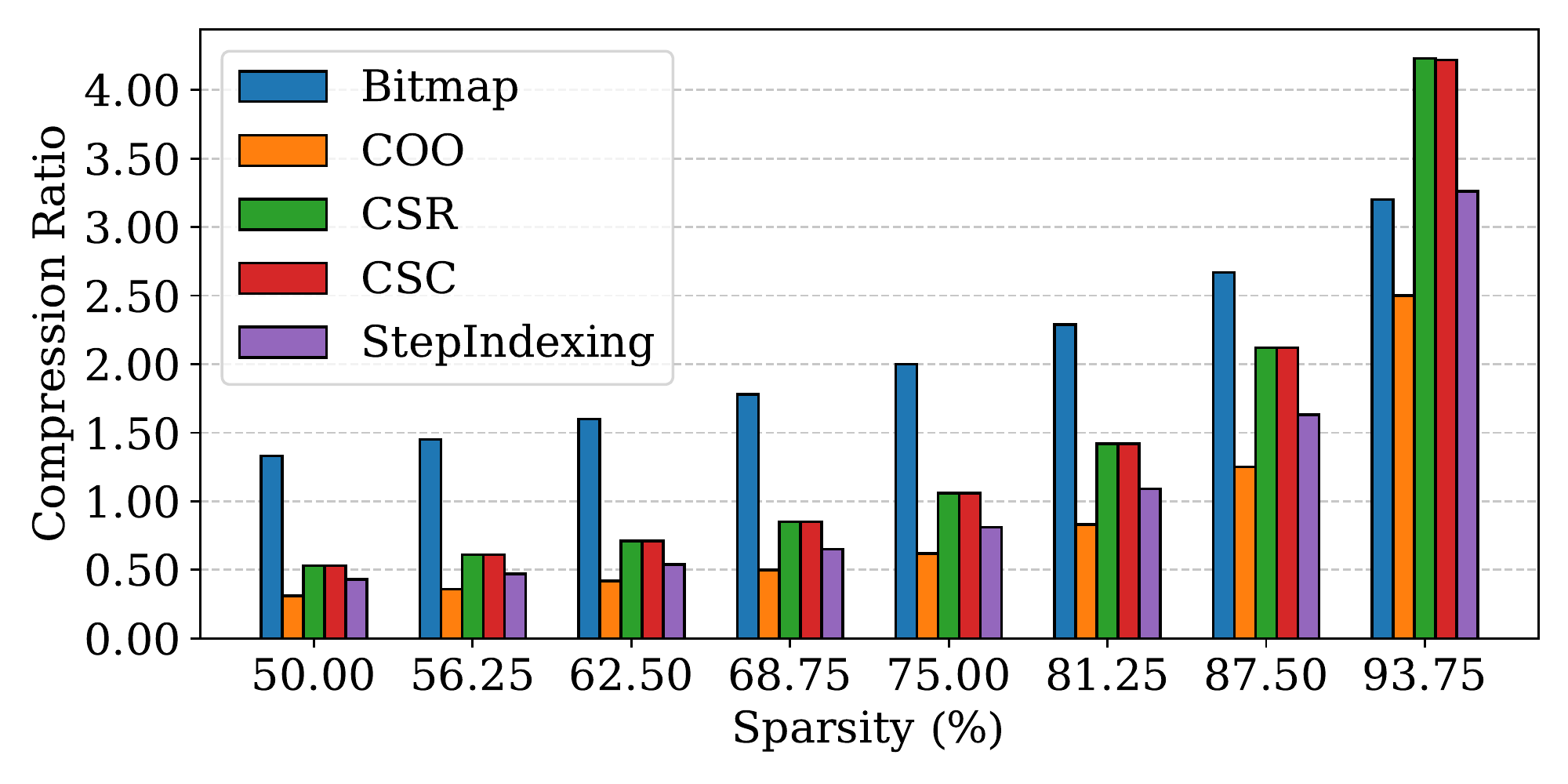}}
    \subfloat[Bitwidth = 8bit]{\includegraphics[width=0.23\textwidth]{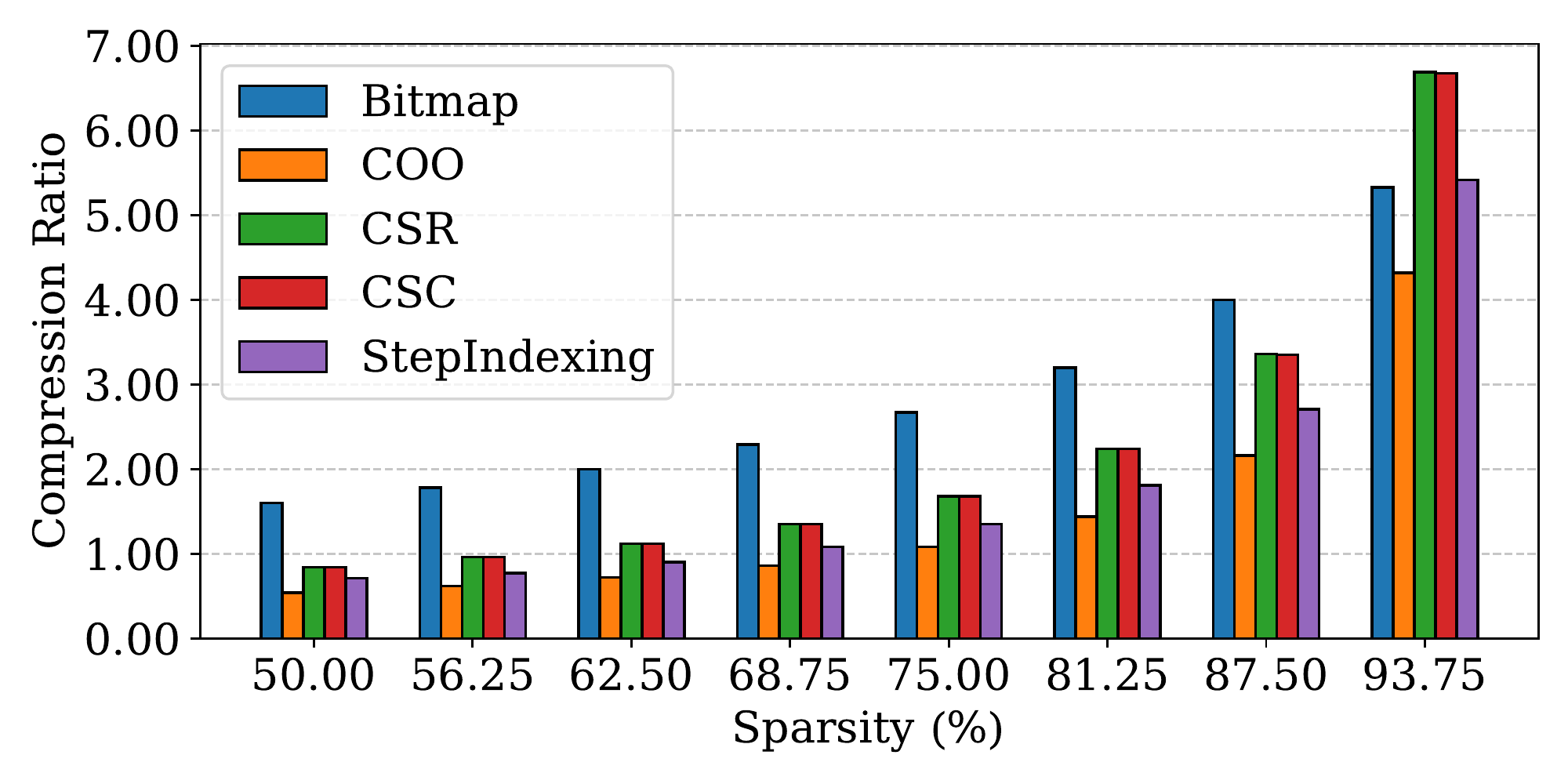}}
    \\
    \subfloat[Bitwidth = 16bit]{\includegraphics[width=0.23\textwidth]{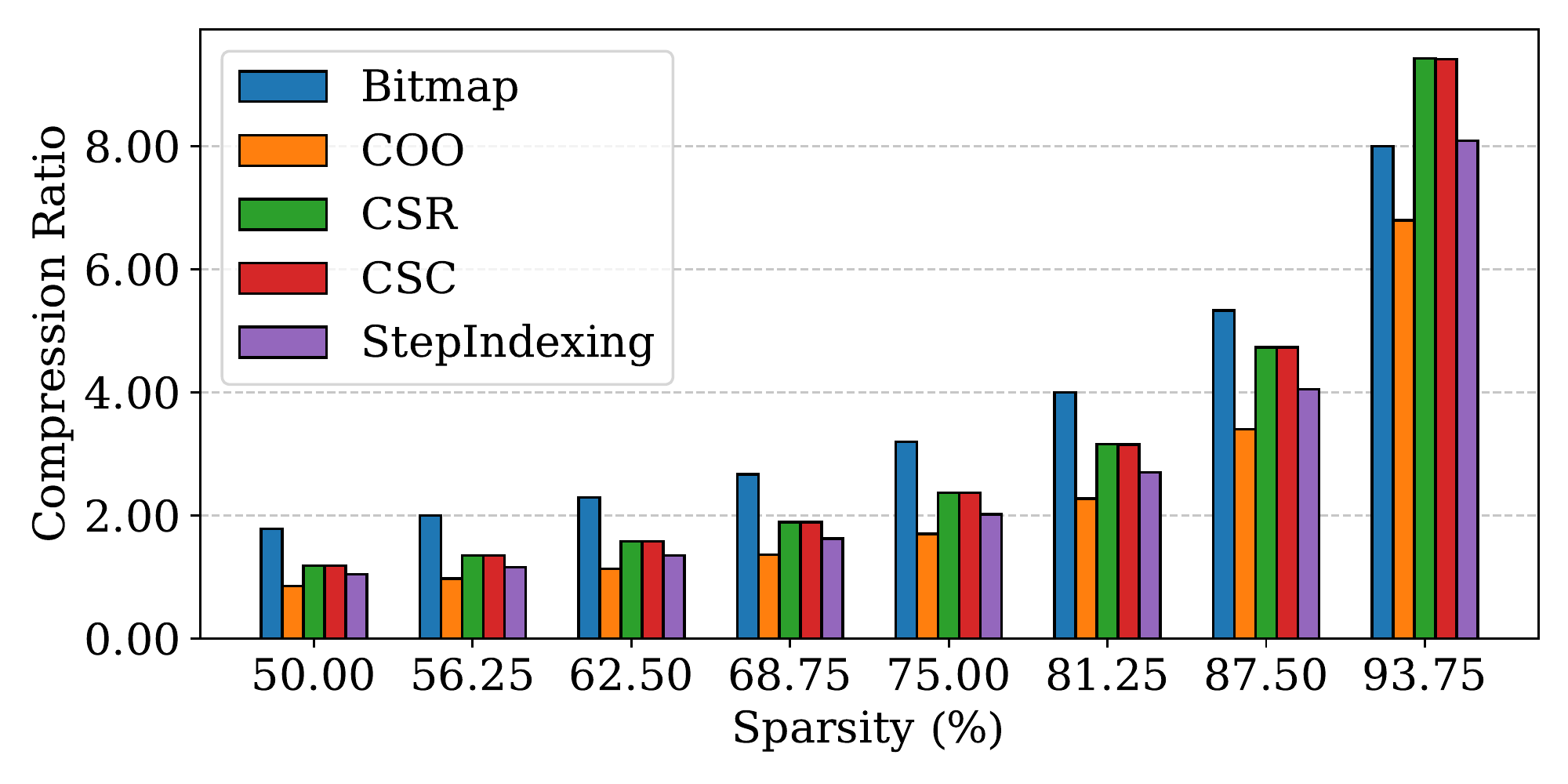}}
    \subfloat[Bitwidth = 32bit]{\includegraphics[width=0.23\textwidth]{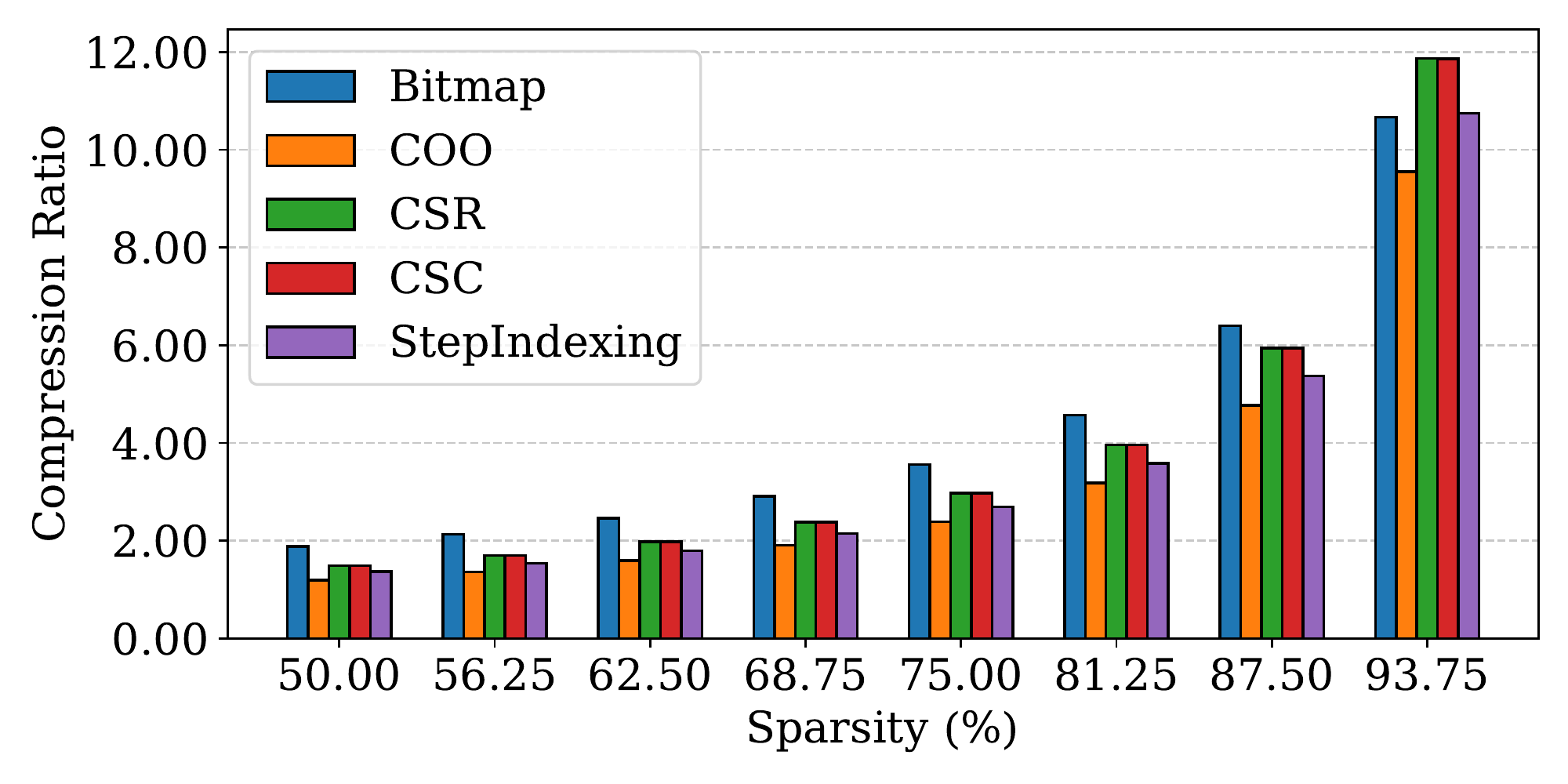}}
    \caption{\RoundOne{Compression ratio on BERT with various sparsity configurations.}}
    \label{fig:res_comp}
\end{figure}

Based on our evaluations of model accuracy with respect to parameter
sparsity, as shown in Fig.~\ref{fig:res_acc}, it is observed that Transformers can hardly achieve a sparsity over $90\%$ without impacting accuracy. It would be more likely practical for \textit{N:M} sparse Transformers with a sparsity ranging from $50\%$ to $87.5\%$.
Next, BERT is taken as an example to evaluate the storage reduction of our compression scheme when using multiple quantized bits under various \textit{N:M} sparsity configurations.
We make a elaborated comparison between our bitmap-based scheme, COO, CSR, CSC, and step indexing \cite{zhang2016cambricon}.
Compared with the other mainstream methods, as shown in Fig.~\ref{fig:res_comp}, our scheme can achieve the highest compression ratio when the model sparsity is varied from $50\%$ to $87.5\%$.
The compression ratio keeps increasing as the model sparsity increases. 
An \textit{N:M} sparse BERT, can achieve a higher compression ratio when quantized in larger bit widths.
BERT with 50.00\% \textit{N:M} sparsity can reach a $1.78\times$ reduction on storage of parameters.
When BERT has a sparsity of 87.50\%, it achieves a significant storage saving, up to $5.33\times$, on parameters.
Our compression scheme can efficiently reduce the storage requirement for \textit{N:M} sparse parameters.
In subsequent hardware evaluations, we uniformly adopt a 16-bit fixed-point representation for Transformers to avoid negative impacts on model accuracy due to quantization.

\subsection{Benchmark Set II: Hardware Resource Consumption}


The second benchmark set verifies hardware consumption of STAs that have not yet deployed Transformer models.

At first, we evaluate the hardware requirements of DMME in common sparse configurations by comparing it against multiple dense computing engines.
For a fair comparison, evaluated computing engines are all synthesized as a $2 \times 2$ unified MatMul computing engine
, and only PEs in these engines are configured into various \textit{N:M} configurations.
Note that those computing engines that make \textit{N} equal to \textit{M}, exclude the non-zero element selectors and merely support dense computing.
It can be regarded as the computing engine in \cite{lu2020hardware} if both \textit{N} and \textit{M} are \textit{1} in the evaluated DMME.
These computing engines are not allowed to be synthesized using DSP blocks, which can be done by setting the property \textit{MAX\_DSP} as zero.
Hence, the utilization of LUTs and FFs from Vivado synthesis reports can be used to measure the consumption of combinational logic and sequential logic for DMMEs, respectively.

\begin{figure}[htb]
    \centering
    \subfloat[Combinational logic]{
        \includegraphics[width=0.23\textwidth]{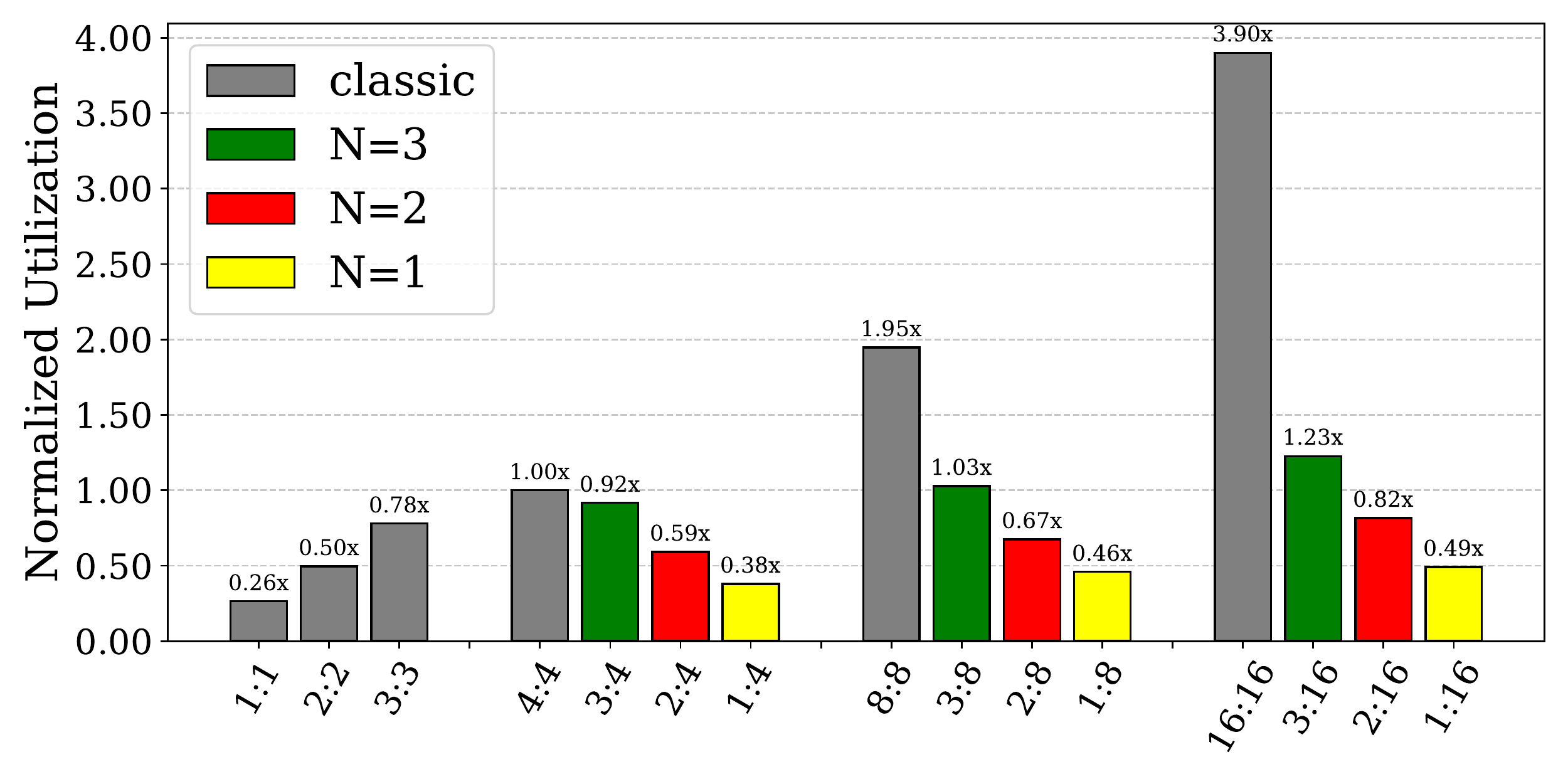}
    }
    \subfloat[Sequential logic]{
        \includegraphics[width=0.23\textwidth]{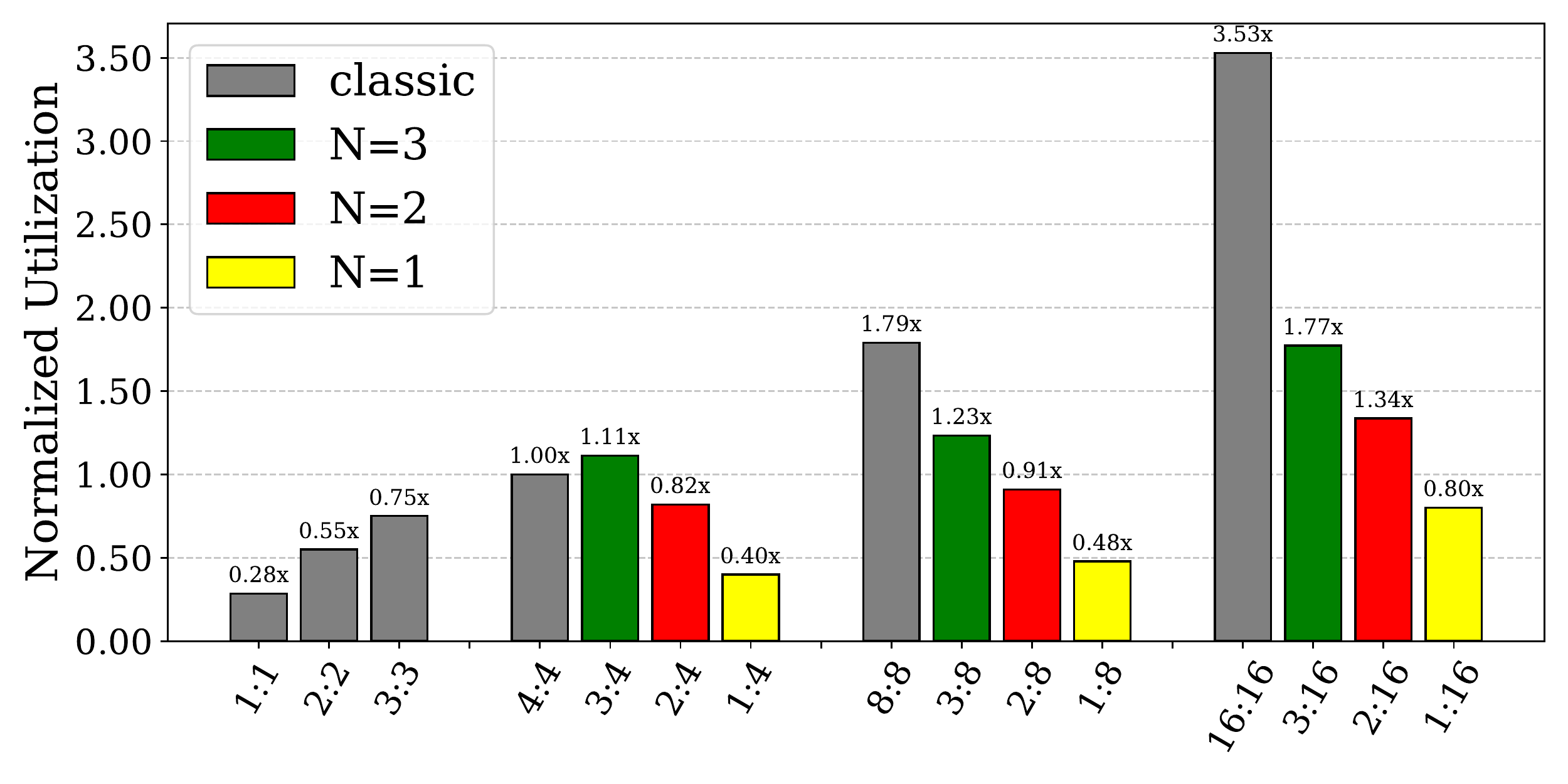}
    }
    \caption{The normalized resource consumption of unified computing engine over the classic computing engine on various scales including (a) combinational logic and (b) sequential logic.}
    \label{fig:res_area}
\end{figure}

Fig.~\ref{fig:res_area} presents the comparison of required hardware resource consumption between DMMEs of various configurations.
For simplicity, all results are normalized to the \textit{4:4} dense baseline computing engine.
Gray bars 
are resource consumption of various computing engines with sole support on dense matrix multiplication.
Green, red, and yellow bars represent hardware utilization of DMMEs, when \textit{N} is set as 1, 2, and 3, respectively.
In Fig.~\ref{fig:res_area}, we can observe hardware resource saved by DMMEs compared to dense baseline computing engine under sparse matrix computing mode.
When $M=16$ and \textit{N} is set to 1, 2, and 3, respectively, DMME, in contrast to \textit{16:16} baseline, obtains saving of combinational logic up to $7.96 \times$, $4.76 \times$, and $3.17 \times$ , while achieving reduction on sequential logic $4.41\times$, $2.63\times$, and $1.99\times$.
According to Fig.~\ref{fig:res_area}, we further evaluate the impact of separately increasing \textit{N} and \textit{M} in DMMEs on hardware resource consumption.
For instance, \textit{3:4} DMME costs $2.42\times$ combinational logic and $2.78\times$ sequential logic of \textit{1:4} DMME.
However, \textit{1:16} DMME merely requires $1.28\times$ combinational logic and $2.00\times$ sequential logic over \textit{1:4} DMME.

We finally evaluate hardware resource consumption of STA on three types of FPGA platforms, including XC7Z020, XC7VX485T, and XCVU13P.
These FPGA platforms are used to represent diverse devices in Fig.~\ref{fig:model_overview}.
Considering the hardware resource and cost on these platforms, we intend to deploy XC7Z020 to the edge and XC7VX485T and XCVU13P on the clouds.
There are many tunable parameters of STAs, especially in DMME, which have great impacts on performance.
In order to determine the specific parameters of STA on each FPGA platforms, we design a cycle-accurate simulator to evaluate actual inference performance based on the given specifications.
STA in XC7Z020 adopts an aggressive \textit{1:8} sparsity since latency is the critical metric on the edge platforms.
However, STAs in both XC7VX485T and XCVU13P configured \textit{N:M} as \textit{2:8} because devices deployed on the clouds concerns latency as well as model accuracy.
Table~\ref{tab:util} shows resource consumption of STAs deployed on three scales of FPGA platforms, namely STA-Tiny, STA-Small, and STA-Large, respectively.
\RoundOne{
The FPGA resource and power breakdown of STA-Small are presented in Table~\ref{tab:breakdown}.
The \textit{N}-parallel MAC (N-MAC) module dominates the DSP consumption of STA since it is the core of computing engine for MAC operations.
The non-zero element selector (NZES) module takes the majority of LUT consumption due to the cost for decoding and index selection.
The routing module occupies most registers of DMME for datapath selection and temporary data storage.
Moreover, the proposed DMME and softmax module occupy $47.29\%$ and $9.42\%$ power consumption, respectively.
}



\begin{table}[htb]
\centering
\caption{FPGA resource utilization}
\label{tab:util}
\resizebox{0.48\textwidth}{!}{%
\begin{tabular}{@{}c|ccccc@{}}
\toprule
Platform                                                            & Frequency & LUT                                                   & FF                                                    & BRAM                                                  & DSP                                                   \\ \midrule
\begin{tabular}[c]{@{}c@{}}1:8 STA-Tiny\\ (XC7Z020)\end{tabular}    & 150MHz  
& \begin{tabular}[c]{@{}c@{}} 21K \\ (40.38\%)\end{tabular} & \begin{tabular}[c]{@{}c@{}} 75K \\ (71.21\%)\end{tabular} & \begin{tabular}[c]{@{}c@{}} 96\\ (68.57\%)\end{tabular} & \begin{tabular}[c]{@{}c@{}} 132\\ (60.00\%)\end{tabular} \\ \midrule
\begin{tabular}[c]{@{}c@{}}2:8 STA-Small\\ (XC7VX485T)\end{tabular} & 200MHz
& \begin{tabular}[c]{@{}c@{}} 116K\\ (38.42\%)\end{tabular} & \begin{tabular}[c]{@{}c@{}} 337K\\ (55.52\%)\end{tabular} & \begin{tabular}[c]{@{}c@{}} 532\\ (51.65\%)\end{tabular} & \begin{tabular}[c]{@{}c@{}} 1,040\\ (37.14\%)\end{tabular} \\ \midrule
\begin{tabular}[c]{@{}c@{}}2:8 STA-Large\\ (XCVU13P)\end{tabular}   & 200MHz
& \begin{tabular}[c]{@{}c@{}} 464K\\ (26.88\%)\end{tabular} & \begin{tabular}[c]{@{}c@{}} 1,321K\\ (38.24\%)\end{tabular} & \begin{tabular}[c]{@{}c@{}} 1,192\\ (44.35\%)\end{tabular} & \begin{tabular}[c]{@{}c@{}} 4,160\\ (33.85\%)\end{tabular} \\ \bottomrule
\end{tabular}%
}
\end{table}

\begin{table}[bht]
\centering
\caption{\RoundOne{FPGA Resource and Power Breakdown of STA-Small}}
\label{tab:breakdown}
\resizebox{0.48\textwidth}{!}{%
\begin{tabular}{@{}cc|ccccc@{}}
\toprule
\multicolumn{2}{c|}{{\color[HTML]{000000} }}                                                        & {\color[HTML]{000000} LUT}                                                              & {\color[HTML]{000000} FF}                                                                & {\color[HTML]{000000} BRAM}                                                             & {\color[HTML]{000000} DSP}                                                               & {\color[HTML]{000000} Power (W)}                                                         \\ \midrule
\multicolumn{1}{c|}{{\color[HTML]{000000} }}                       & {\color[HTML]{000000} N-MAC}   & {\color[HTML]{000000} \begin{tabular}[c]{@{}c@{}}16K\\ (13.79\%)\end{tabular}}          & {\color[HTML]{000000} \begin{tabular}[c]{@{}c@{}}68K\\ (20.18\%)\end{tabular}}           & {\color[HTML]{000000} \textbf{-}}                                                       & {\color[HTML]{000000} \textbf{\begin{tabular}[c]{@{}c@{}}1024\\ (98.46\%)\end{tabular}}} & {\color[HTML]{000000} \begin{tabular}[c]{@{}c@{}}2.78\\ (28.17\%)\end{tabular}}          \\ \cmidrule(l){2-7} 
\multicolumn{1}{c|}{{\color[HTML]{000000} }}                       & {\color[HTML]{000000} NZES}    & {\color[HTML]{000000} \textbf{\begin{tabular}[c]{@{}c@{}}66K\\ (56.90\%)\end{tabular}}} & {\color[HTML]{000000} \begin{tabular}[c]{@{}c@{}}68K\\ (20.18\%)\end{tabular}}           & {\color[HTML]{000000} -}                                                                & {\color[HTML]{000000} -}                                                                 & {\color[HTML]{000000} \begin{tabular}[c]{@{}c@{}}0.83\\ (8.41\%)\end{tabular}}           \\ \cmidrule(l){2-7} 
\multicolumn{1}{c|}{\multirow{-6}{*}{{\color[HTML]{000000} DMME}}} & {\color[HTML]{000000} Routing} & {\color[HTML]{000000} \begin{tabular}[c]{@{}c@{}}9K\\ (7.76\%)\end{tabular}}            & {\color[HTML]{000000} \textbf{\begin{tabular}[c]{@{}c@{}}180K\\ (53.41\%)\end{tabular}}} & {\color[HTML]{000000} -}                                                                & {\color[HTML]{000000} -}                                                                 & {\color[HTML]{000000} \begin{tabular}[c]{@{}c@{}}1.06\\ (10.74\%)\end{tabular}}          \\ \midrule
\multicolumn{2}{c|}{{\color[HTML]{000000} Softmax}}                                                 & {\color[HTML]{000000} \begin{tabular}[c]{@{}c@{}}13K\\ (11.21\%)\end{tabular}}          & {\color[HTML]{000000} \begin{tabular}[c]{@{}c@{}}8K\\ (2.37\%)\end{tabular}}             & {\color[HTML]{000000} \begin{tabular}[c]{@{}c@{}}16\\ (3.00\%)\end{tabular}}            & {\color[HTML]{000000} \begin{tabular}[c]{@{}c@{}}16\\ (1.54\%)\end{tabular}}             & {\color[HTML]{000000} \begin{tabular}[c]{@{}c@{}}0.93\\ (9.42\%)\end{tabular}}           \\ \midrule
\multicolumn{2}{c|}{{\color[HTML]{000000} Others}}                                                  & {\color[HTML]{000000} \begin{tabular}[c]{@{}c@{}}12K\\ (10.34\%)\end{tabular}}          & {\color[HTML]{000000} \begin{tabular}[c]{@{}c@{}}13K\\ (3.86\%)\end{tabular}}            & {\color[HTML]{000000} \textbf{\begin{tabular}[c]{@{}c@{}}516\\ (97.00\%)\end{tabular}}} & {\color[HTML]{000000} \textbf{-}}                                                        & {\color[HTML]{000000} \textbf{\begin{tabular}[c]{@{}c@{}}4.27\\ (43.26\%)\end{tabular}}} 
\\ \midrule
\multicolumn{2}{c|}{{\color[HTML]{000000} Total}}                                                   & {\color[HTML]{000000} \begin{tabular}[c]{@{}c@{}}116K\\ (100.00\%)\end{tabular}}        & {\color[HTML]{000000} \begin{tabular}[c]{@{}c@{}}337K\\ (100.00\%)\end{tabular}}         & {\color[HTML]{000000} \begin{tabular}[c]{@{}c@{}}532\\ (100.00\%)\end{tabular}}         & {\color[HTML]{000000} \begin{tabular}[c]{@{}c@{}}1040\\ (100.00\%)\end{tabular}}         & {\color[HTML]{000000} \begin{tabular}[c]{@{}c@{}}9.87\\ (100.00\%)\end{tabular}}         
\\ \bottomrule
\end{tabular}%
}
\end{table}

\subsection{Benchmark Set III: Overall System Evaluation}

The third benchmark set is used to evaluate performance when deploying various Transformer-based models on STAs.

\begin{figure}[bht]
    \centering
    \subfloat[]{
        \includegraphics[width=0.48\textwidth]{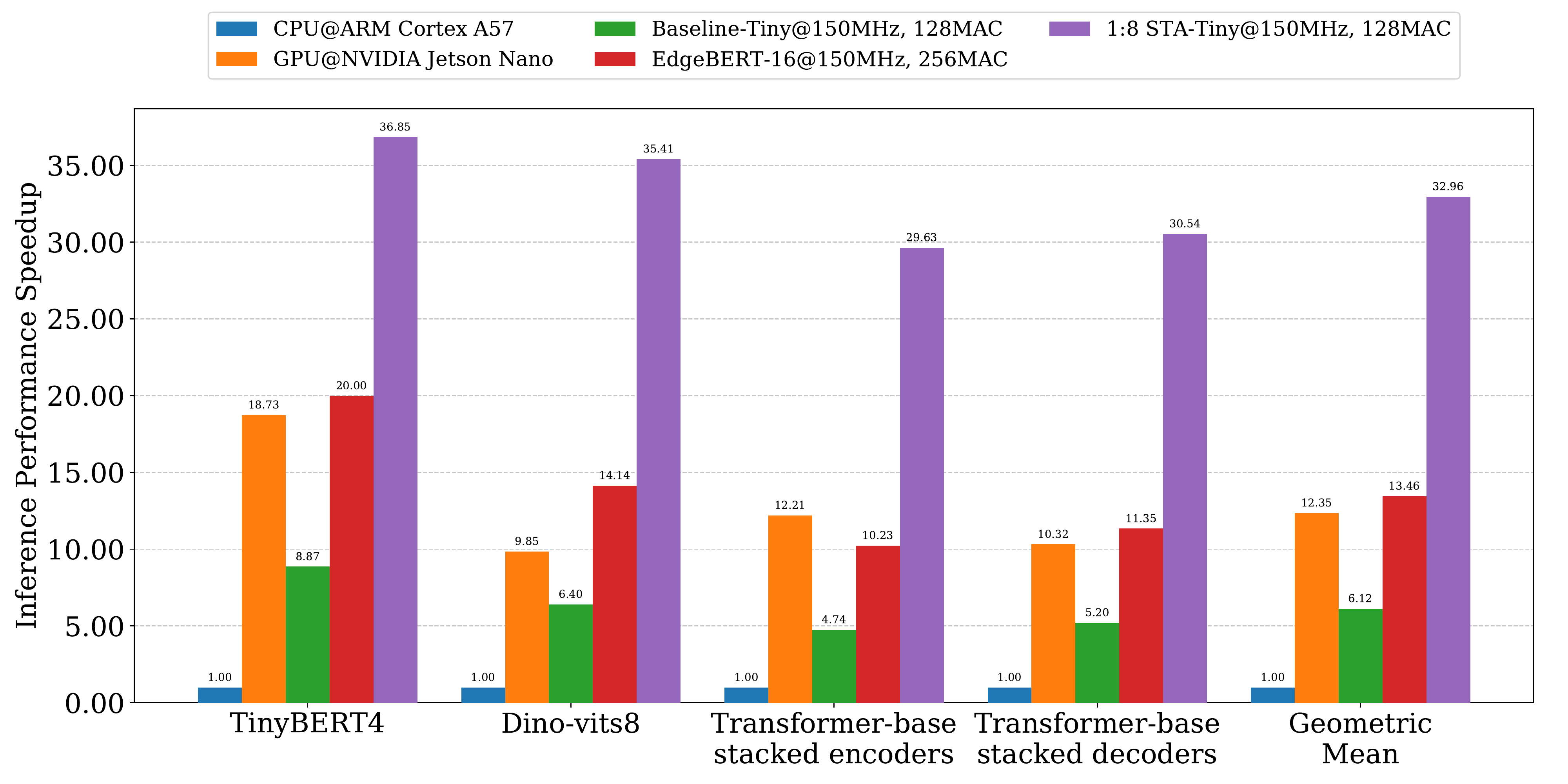}
        \label{fig:res_edge_latency}
    }
    \\
    \subfloat[]{
        \includegraphics[width=0.48\textwidth]{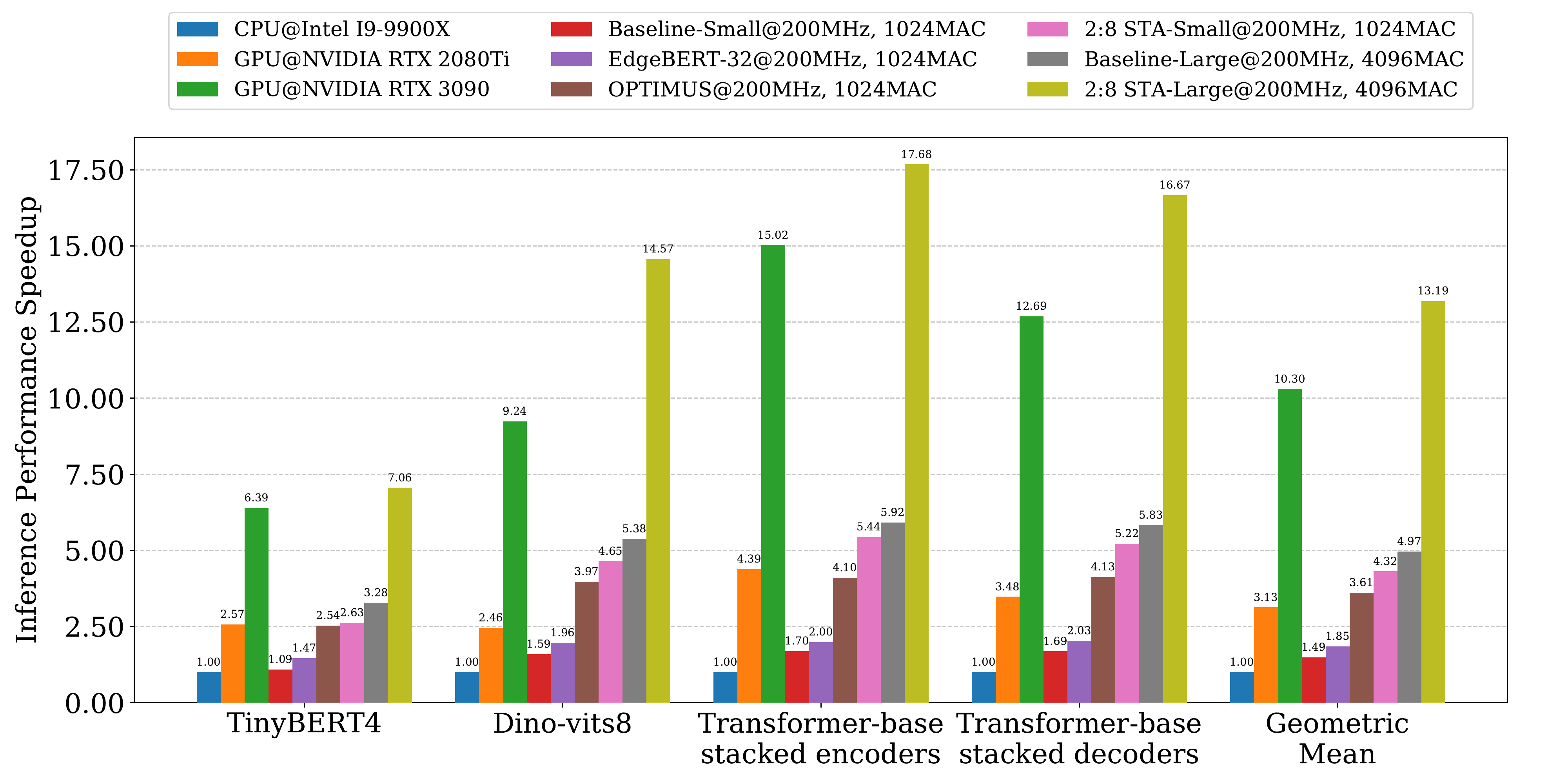}
        \label{fig:res_cloud_latency}
    }
    \caption{\RoundOne{The processing time of Transformer-based models on various (a) edge platforms and (b) cloud platforms.}}
    \label{fig:res_latency}
\end{figure}

Firstly, we study the inference speedup of STAs by contrast with CPUs, GPUs, and the prior dedicated accelerators.
The selected models to be deployed are composed of TinyBERT \cite{jiao2020tinybert}, Dino \cite{caron2021emerging}, the classic Transformer model \cite{vaswani2017attention}.
Here we consider single-batch processing time in all the MHA and FFN ResBlocks of these Transformer-based models. 
For cross-platform comparison, the hardware setup is as follows to execute the Transformer inference tasks.
The CPU results are measured using an ARM Cortex A57 and an Intel i9-9900X.
The former commonly appeared in mobile devices for edge applications, while the latter is a high-end CPU product for deploying cloud applications.
\RoundOne{The GPU results are measured using an NVIDIA Jeston Nano, an embedded GPU product for edge applications, an NVIDIA RTX 2080Ti, and an NVIDIA RTX 3090 capable of 2:4 sparse acceleration.}
Following the comparison method in \cite{lu2021sanger}, we apply \cite{lu2020hardware} as baselines on FPGA platforms by scaling the size of its computing engine.
\RoundOne{Two existing sparse accelerators for Transformers, OPTIMUS \cite{park2020optimus} and EdgeBERT \cite{tambe2021edgebert}, are evaluated as well for a more comprehensive comparison of STA.}
Fig.~\ref{fig:res_latency} shows the idealized performance speedup of different hardware platforms, where edge and cloud platforms normalized to the ARM Cortex A57 and Intel i9-9900X, respectively.

\begin{figure} [tbp] 
	\centering
	\includegraphics[width=0.46\textwidth]{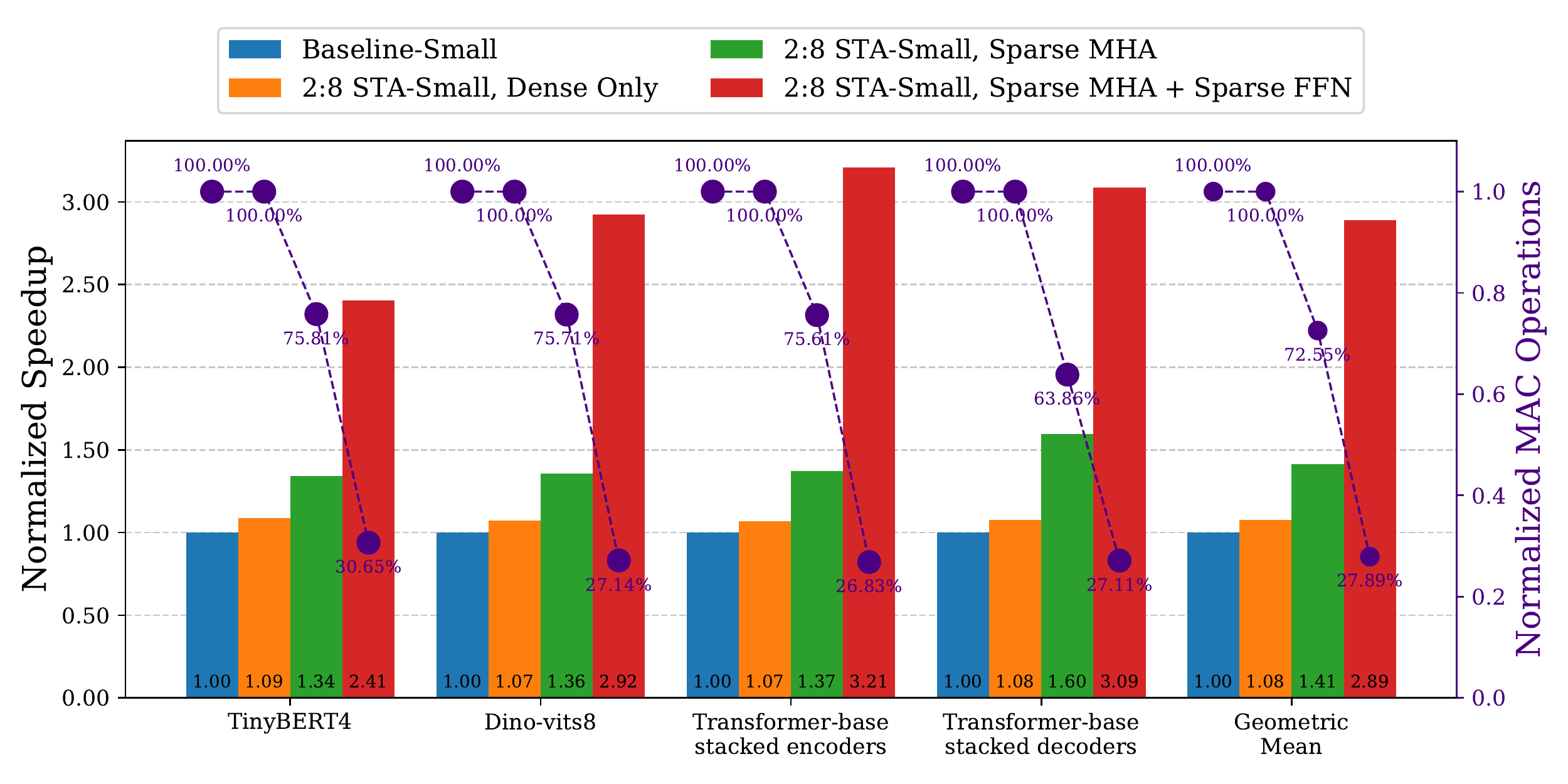}
	\caption{\RoundOne{Performance and MAC operation breakdown of STA-Small.}}
	\label{fig:sta_small_perf}
\end{figure}

As shown in Fig.~\ref{fig:res_latency}~(a), among all edge platforms, STA-Tiny achieves a geometric mean increase of $32.96\times$, $2.69\times$, and $5.38 \times$ over CPU, GPU, and the FPGA baseline, respectively.
Fig.~\ref{fig:res_latency}~(b) presents the performance comparison between various cloud platforms.
\RoundOne{
For fair comparison, Baseline-Small (red), EdgeBERT (purple), OPTIMUS (brown), and \textit{2:8} STA-Small (pink) in Fig.~\ref{fig:res_latency}~(b) are evaluated under the same number of MAC units and clock frequency.
}
\RoundOne{
\begin{itemize}
    \item \textbf{STA v.s. Baseline-Small} \cite{lu2020hardware}: 
    \textit{2:8} STA-Small achieves $2.89\times$ speedup on average over Baseline-Small, which enables dense Transformer acceleration using a large 2D systolic array.
    It suffers low utilization of MAC units due to inflexible mapping scheme and large skew latency for systolic-arranged data. 
    STA achieves significant performance improvement by: 1) innovation of DMME from the architectural aspects, and 2) reduction on MAC operations of \textit{N:M} sparse Transformers from the algorithmic aspects.
    We further present a performance breakdown of STA compared to Baseline-Small.
    As shown in Fig.~\ref{fig:sta_small_perf}, the architectural innovation for DMME can achieve $1.08\times$ better performance improvement, and there is $2.68\times$ speedup on top of the architectural innovation by efficiently enabling \textit{2:8} sparse acceleration.
    \item \textbf{STA v.s. EdgeBERT} \cite{tambe2021edgebert}:
    \textit{2:8} STA-Small achieves $2.33\times$ speedup on average over EdgeBERT-32, which is an energy-optimized Transformer accelerator exploiting unstructured sparsity.
    When performing sparse MAC operations, processing units of EdgeBERT skip the zero input value through a gating strategy, which can significantly reduce energy consumption, but offer little benefit to latency.
    Compared to EdgeBERT, STA performs \textit{N:M} sparse MAC operations by choosing non-zero inputs, which can both reduce energy and optimize latency.
    \item \textbf{STA v.s. OPTIMUS} \cite{park2020optimus}:
    \textit{2:8} STA-Small has a $1.20\times$ better performance on average over OPTIMUS, which is a high performance sparse accelerator for Transformers exploiting unstructured sparsity in weight parameters. 
    When performing sparse MAC operations, OPTIMUS can hardly achieve high MAC utilization due to load imbalance and input load miss.
    STA can effectively overcome these two problem suffered by OPTIMUS.
    STA gets rid of load imbalance by arranging each MAC in DMME to perform operations in a balanced \textit{N:M} group.
    In addition, STA loads a series of input \textit{N:M} groups at each cycle, and utilizes these elements multiple times in a systolic manner, which effectively addresses input load miss compared to OPTIMUS.
\end{itemize}
}


\begin{table*}[ht]
\centering
\caption{\RoundOne{Comparison of STAs with previous works and commercial products}}
\label{tab:perf_cmp}
\resizebox{0.90\textwidth}{!}{%
\begin{tabular}{@{}c|cccccccccc@{}}
\toprule
\multirow{3}{*}{Platform}   & CPU                                            & \multicolumn{3}{|c}{GPU}                                                                                                                               & \multicolumn{6}{|c}{FPGA}                                                                                                                                                                                                                                                                                                                                 \\ \cmidrule(l){2-11} 
                            & \multicolumn{1}{c|}{\multirow{2}{*}{i9-9900X}} & \multirow{2}{*}{Jetson Nano} & \multirow{2}{*}{RTX 2080 Ti} & \multicolumn{1}{c|}{\multirow{2}{*}{\RoundOne{RTX 3090}}}                                          & \multirow{2}{*}{\begin{tabular}[c]{@{}c@{}}SOCC'20\\ \cite{lu2020hardware}\end{tabular}} & \multirow{2}{*}{\begin{tabular}[c]{@{}c@{}}ISLPED'20\\ \cite{li2020ftrans}\end{tabular}} & \multicolumn{1}{c|}{\multirow{2}{*}{\begin{tabular}[c]{@{}c@{}}ISQED'21\\ \cite{peng2021accelerating}\end{tabular}}} & \multicolumn{3}{c}{Our work}                \\ \cmidrule(l){9-11} 
                            & \multicolumn{1}{c|}{}                          &                              &                              & \multicolumn{1}{c|}{}                                                                   &                                                                                          &                                                                                          & \multicolumn{1}{c|}{}                                                                                                & STA-Tiny     & STA-Small     & STA-Large    \\ \midrule
Chip                        & \multicolumn{1}{c|}{Skylake}                   & Tegra X1                     & TU102                        & \multicolumn{1}{c|}{\RoundOne{GA102}}                                                              & XCVU13P                                                                                  & XCVU9P                                                                                   & \multicolumn{1}{c|}{XCU200}                                                                                          & XC7Z020      & XC7VX485T     & XCVU13P      \\ \midrule
Technology                  & \multicolumn{1}{c|}{14 nm}                     & 20 nm                        & 12 nm                        & \multicolumn{1}{c|}{\RoundOne{8 nm}}                                                               & 16 nm                                                                                    & 16 nm                                                                                    & \multicolumn{1}{c|}{16 nm}                                                                                           & 28 nm        & 28 nm         & 16 nm        \\ \midrule
Frequency                   & \multicolumn{1}{c|}{3.50 GHz}                  & 640 MHz                      & 1.35 GHz                     & \multicolumn{1}{c|}{\RoundOne{1.70 GHz}}                                                           & 200 MHz                                                                                  & -                                                                                        & \multicolumn{1}{c|}{-}                                                                                               & 150 MHz      & 200 MHz       & 200 MHz      \\ \midrule
Methods                     & \multicolumn{1}{c|}{-}                         & -                            & -                            & \multicolumn{1}{c|}{\begin{tabular}[c]{@{}c@{}}\RoundOne{2:4 group-based} \\ \RoundOne{pruning}\end{tabular}} & \begin{tabular}[c]{@{}c@{}}Low-bit\\ quantization\end{tabular}                           & \begin{tabular}[c]{@{}c@{}}Block-circulant\\ matrix with FFT\end{tabular}                & \multicolumn{1}{c|}{\begin{tabular}[c]{@{}c@{}}Block-based\\ pruning\end{tabular}}                                   & \multicolumn{3}{c}{N:M group-based pruning} \\ \midrule
\# MAC units                 & \multicolumn{1}{c|}{-}                         & -                            & -                            & \multicolumn{1}{c|}{\RoundOne{-}}                                                                  & 4096                                                                                     & $\sim$ 5647                                                                              & \multicolumn{1}{c|}{$\sim$ 3368}                                                                                     & 128          & 1024          & 4096         \\ \midrule
Bit Precision               & \multicolumn{1}{c|}{FP-32}                     & FP-32                        & FP-32                        & \multicolumn{1}{c|}{\RoundOne{FP-32}}                                                              & FIX-8                                                                                    & FIX-16                                                                                   & \multicolumn{1}{c|}{-}                                                                                               & \multicolumn{3}{c}{FIX-16}                  \\ \midrule
Test Network                & \multicolumn{10}{c}{Shallow Transformer}                                                                                                                                                                                                                                                                                                                                                                                                                                                                                                                          \\ \midrule
Latency (ms)                & \multicolumn{1}{c|}{2.17}                      & 16.24                        & 1.70                         & \multicolumn{1}{c|}{\RoundOne{0.46}}                                                               & 0.30                                                                                     & 2.94                                                                                     & \multicolumn{1}{c|}{0.32}                                                                                            & 2.01         & 0.42          & 0.15         \\ \midrule
Batch-1 Throughput (GOP/s)  & \multicolumn{1}{c|}{101.38}                    & 13.55                        & 129.41                       & \multicolumn{1}{c|}{\RoundOne{478.26}}                                                             & 733.33                                                                                   & 75.34                                                                                    & \multicolumn{1}{c|}{687.50}                                                                                          & 109.45       & 523.81        & 1466.67      \\ \midrule
Power (W)                   & \multicolumn{1}{c|}{165.00}                    & 7.56                         & 250.00                       & \multicolumn{1}{c|}{\RoundOne{350.00}}                                                             & 16.70                                                                                    & 22.45                                                                                    & \multicolumn{1}{c|}{-}                                                                                               & 2.71         & 9.87          & 26.59        \\ \midrule
Energy Efficiency (GOP/J)   & \multicolumn{1}{c|}{0.61}                      & 1.79                         & 0.52                         & \multicolumn{1}{c|}{\RoundOne{1.37}}                                                               & 43.91                                                                                    & 3.35                                                                                     & \multicolumn{1}{c|}{-}                                                                                               & 40.39        & 53.07         & 55.16        \\ \midrule
MAC Efficiency (GOP/s/unit) & \multicolumn{1}{c|}{-}                         & -                            & -                            & \multicolumn{1}{c|}{\RoundOne{-}}                                                                  & 0.18                                                                                     & $\sim$ 0.01                                                                              & \multicolumn{1}{c|}{$\sim$ 0.20}                                                                                                          & 0.86         & 0.51          & 0.36         \\ \bottomrule
\end{tabular}%
}
\end{table*}

Finally, we compare STA with other previous FPGA-based works and commercial CPU and GPU products.
Table~\ref{tab:perf_cmp} presents a fair performance comparison without batching on various platforms.
Prior FPGA-based works for accelerating Transformers, include \cite{lu2020hardware}, \cite{li2020ftrans}, and \cite{peng2021accelerating}.
The dedicated accelerator in \cite{lu2020hardware}, equipped with a large 2D systolic array for dense operation, is the pioneer work for Transformer acceleration.
FTRANS \cite{li2020ftrans}, is another recent specialized accelerator for Transformers, which exploits the speedup potential of block-circulant weight representations.
In \cite{peng2021accelerating}, the proposed accelerator utilizes coarse-grained block-based sparsity to speedup Transformer inference.
The shallow Transformer used in \cite{li2020ftrans} and \cite{peng2021accelerating} is applied as a benchmark for a fair evaluation.
The comparison is benchmarked on CPU, GPUs, prior cutting-edge FPGA solutions, and STA on various FPGA platforms.
We evaluate these designs in terms of latency, throughput, power, energy efficiency, and MAC efficiency.
All of them are key metrics for a computing system.

As shown in Table~\ref{tab:perf_cmp}, STA-Tiny far outperforms the embedded GPU, Jetson Nano, and the high-end CPU, i9-9900X, in all evaluated metrics.
STA-Small surpasses the CPU and GPU platforms in all metrics.
Moreover, STA-Small is close to \cite{lu2020hardware} and \cite{peng2021accelerating} in terms of latency and throughput, while using a relatively small number of MACs compared to them.
The energy efficiency and MAC efficiency of STA-Small are also superior to all previous FPGA-based works.
Compared to previous FPGA solutions, STA-Large achieves $2.00 \sim 19.47 \times$ throughput improvement, achieves $1.26 \sim 16.47 \times$ energy efficiency improvement, and $1.80 \sim 36.00\times$ MAC efficiency gain, respectively.
The performance gain of STA comes from optimizations from two levels.
At the algorithm level, we carefully exploit the potential of \textit{N:M} sparsity pattern, which can significantly reduce the computational cost of Transformer-based models.
At the hardware level, STA can efficiently handle \textit{N:M} sparse parameters, which significantly improves the utilization of computing units.
In addition, our deployment framework, taking STA-Tiny, STA-Small, and STA-Large as examples, can realize flexible hardware generation for Transformers.
The proposed framework can flexibly and efficiently meet the requirements for deploying Transformer-based models on various FPGA devices.




\section{Conclusion}
In this paper, we present a flexible, agile, and efficient framework for deploying \textit{N:M} sparse Transformers, which is benefited from both algorithm and hardware optimizations, making it practical to significantly accelerate Transformer-based models on diverse FPGA devices.
At the algorithm level, we propose a sparsity inheritance mechanism and a inherited dynamic pruning (IDP) method to obtain a series of \textit{N:M} sparse Transformers with high accuracy. 
A further proposed compression scheme greatly reduces the storage requirements of models.
At the hardware level, we present a flexible and efficient architecture, namely STA, to accelerate \textit{N:M} sparse Transformers.
STA is composed of a computing core, DMME, that unifies both sparse and dense intensive matrix multiplications in \textit{N:M} sparse Transformers, and a scalable softmax module, which eliminates intermediate off-chip data accesses.
The experimental results show that \textit{N:M} sparse Transformers generated by IDP achieves an average of $6.7\%$ improvement in accuracy over the state-of-the-art methods.
STA implementation significantly outperforms CPU, GPU, and prior FPGA-based Transformer accelerators in terms of latency, throughput, energy efficiency, and MAC efficiency, showing its significant potential in applications using Transformer-based models.

\section*{Acknowledgment}

We would like to sincerely thank our reviewers for their valuable feedback.

\ifCLASSOPTIONcaptionsoff
  \newpage
\fi



\bibliographystyle{IEEEtran}
\bibliography{ref.bib}

\begin{thebibliography}{10}
\providecommand{\url}[1]{#1}
\csname url@samestyle\endcsname
\providecommand{\newblock}{\relax}
\providecommand{\bibinfo}[2]{#2}
\providecommand{\BIBentrySTDinterwordspacing}{\spaceskip=0pt\relax}
\providecommand{\BIBentryALTinterwordstretchfactor}{4}
\providecommand{\BIBentryALTinterwordspacing}{\spaceskip=\fontdimen2\font plus
\BIBentryALTinterwordstretchfactor\fontdimen3\font minus
  \fontdimen4\font\relax}
\providecommand{\BIBforeignlanguage}[2]{{%
\expandafter\ifx\csname l@#1\endcsname\relax
\typeout{** WARNING: IEEEtran.bst: No hyphenation pattern has been}%
\typeout{** loaded for the language `#1'. Using the pattern for}%
\typeout{** the default language instead.}%
\else
\language=\csname l@#1\endcsname
\fi
#2}}
\providecommand{\BIBdecl}{\relax}
\BIBdecl

\bibitem{tay2020efficient}
Y.~Tay, M.~Dehghani, D.~Bahri, and D.~Metzler, ``{Efficient Transformers: A
  Survey},'' \emph{arXiv preprint arXiv:2009.06732}, 2020.

\bibitem{song2020alignment}
K.~Song, K.~Wang, H.~Yu, Y.~Zhang, Z.~Huang, W.~Luo, X.~Duan, and M.~Zhang,
  ``{Alignment-Enhanced Transformer for Constraining NMT with Pre-specified
  Translations},'' in \emph{Proceedings of the AAAI Conference on Artificial
  Intelligence (AAAI)}, vol.~34, no.~05, 2020, pp. 8886--8893.

\bibitem{kenton2019bert}
J.~D. M.-W.~C. Kenton and L.~K. Toutanova, ``{BERT: Pre-training of Deep
  Bidirectional Transformers for Language Understanding},'' in
  \emph{Proceedings of the Conference of the North American Chapter of the
  Association for Computational Linguistics – Human Language Technologies
  (NAACL-HLT)}, 2019, pp. 4171--4186.

\bibitem{dosovitskiy2020image}
A.~Dosovitskiy, L.~Beyer, A.~Kolesnikov, D.~Weissenborn, X.~Zhai,
  T.~Unterthiner, M.~Dehghani, M.~Minderer, G.~Heigold, S.~Gelly, J.~Uszkoreit,
  and N.~Houlsby, ``{An Image is Worth 16x16 Words: Transformers for Image
  Recognition at Scale},'' in \emph{International Conference on Learning
  Representations (ICLR)}, 2021.

\bibitem{park2020optimus}
J.~Park, H.~Yoon, D.~Ahn, J.~Choi, and J.-J. Kim, ``{{OPTIMUS}: OPTImized
  Matrix MUltiplication Structure for Transformer Neural Network
  Accelerator},'' in \emph{Proceedings of Machine Learning and Systems
  (MLSys)}, 2020.

\bibitem{tambe2021edgebert}
T.~Tambe, C.~Hooper, L.~Pentecost, T.~Jia, E.-Y. Yang, M.~Donato, V.~Sanh,
  P.~N. Whatmough, A.~M. Rush, D.~Brooks, and G.-Y. Wei, ``{EdgeBERT:
  Sentence-Level Energy Optimizations for Latency-Aware Multi-Task NLP
  Inference},'' in \emph{Proceedings of the 54th Annual IEEE/ACM International
  Symposium on Microarchitecture (MICRO)}, 2021.

\bibitem{zhou2021learning}
A.~Zhou, Y.~Ma, J.~Zhu, J.~Liu, Z.~Zhang, K.~Yuan, W.~Sun, and H.~Li,
  ``{Learning {N:M} Fine-grained Structured Sparse Neural Networks From
  Scratch},'' in \emph{International Conference on Learning Representations
  (ICLR)}, 2021.

\bibitem{sun2021dominosearch}
W.~Sun, A.~Zhou, S.~Stuijk, R.~Wijnhoven, A.~O. Nelson, H.~Corporaal
  \emph{et~al.}, ``{DominoSearch: Find Layer-wise Fine-grained N: M Sparse
  Schemes from Dense Neural Networks},'' \emph{Advances in Neural Information
  Processing Systems (NeurIPS)}, vol.~34, 2021.

\bibitem{mishra2021accelerating}
A.~Mishra, J.~A. Latorre, J.~Pool, D.~Stosic, D.~Stosic, G.~Venkatesh, C.~Yu,
  and P.~Micikevicius, ``{Accelerating Sparse Deep Neural Networks},''
  \emph{arXiv preprint arXiv:2104.08378}, 2021.

\bibitem{lu2020hardware}
S.~Lu, M.~Wang, S.~Liang, J.~Lin, and Z.~Wang, ``{Hardware Accelerator for
  Multi-Head Attention and Position-Wise Feed-Forward in the Transformer},'' in
  \emph{2020 IEEE 33rd International System-on-Chip Conference (SOCC)}, 2020.

\bibitem{vaswani2017attention}
A.~Vaswani, N.~Shazeer, N.~Parmar, J.~Uszkoreit, L.~Jones, A.~N. Gomez,
  {\L}.~Kaiser, and I.~Polosukhin, ``{Attention Is All You Need},'' in
  \emph{Proceedings of the 31st International Conference on Neural Information
  Processing Systems (NeurIPS)}, 2017.

\bibitem{wu2021swm}
D.~Wu, X.~Fan, W.~Cao, and L.~Wang, ``{SWM: A High-Performance Sparse-Winograd
  Matrix Multiplication CNN Accelerator},'' \emph{IEEE Transactions on Very
  Large Scale Integration (VLSI) Systems}, vol.~29, no.~5, pp. 936--949, 2021.

\bibitem{colleman2021high}
S.~Colleman and M.~Verhelst, ``{High-Utilization, High-Flexibility Depth-First
  CNN Coprocessor for Image Pixel Processing on FPGA},'' \emph{IEEE
  Transactions on Very Large Scale Integration (VLSI) Systems}, vol.~29, no.~3,
  pp. 461--471, 2021.

\bibitem{yantir2021imca}
H.~E. Yant{\i}r, A.~M. Eltawil, and K.~N. Salama, ``{IMCA: An Efficient
  In-Memory Convolution Accelerator},'' \emph{IEEE Transactions on Very Large
  Scale Integration (VLSI) Systems}, vol.~29, no.~3, pp. 447--460, 2021.

\bibitem{yin2019vesti}
S.~Yin, Z.~Jiang, M.~Kim, T.~Gupta, M.~Seok, and J.-S. Seo, ``{Vesti:
  Energy-Efficient In-Memory Computing Accelerator for Deep Neural Networks},''
  \emph{IEEE Transactions on Very Large Scale Integration (VLSI) Systems},
  vol.~28, no.~1, pp. 48--61, 2019.

\bibitem{paulin2021rnn}
G.~Paulin, R.~Andri, F.~Conti, and L.~Benini, ``{RNN-Based Radio Resource
  Management on Multicore RISC-V Accelerator Architectures},'' \emph{IEEE
  Transactions on Very Large Scale Integration (VLSI) Systems}, vol.~29, no.~9,
  pp. 1624--1637, 2021.

\bibitem{fang2021accelerating}
C.~Fang, L.~He, H.~Wang, J.~Wei, and Z.~Wang, ``{Accelerating 3D Convolutional
  Neural Networks Using 3D Fast Fourier Transform},'' in \emph{2021 IEEE
  International Symposium on Circuits and Systems (ISCAS)}.\hskip 1em plus
  0.5em minus 0.4em\relax IEEE, 2021, pp. 1--5.

\bibitem{yu2020uni}
Y.~Yu, T.~Zhao, M.~Wang, K.~Wang, and L.~He, ``{Uni-OPU: An FPGA-based Uniform
  Accelerator for Convolutional and Transposed Convolutional Networks},''
  \emph{IEEE transactions on very large scale integration (VLSI) systems},
  vol.~28, no.~7, pp. 1545--1556, 2020.

\bibitem{zhu2020efficient}
C.~Zhu, K.~Huang, S.~Yang, Z.~Zhu, H.~Zhang, and H.~Shen, ``{An Efficient
  Hardware Accelerator for Structured Sparse Convolutional Neural Networks on
  FPGAs},'' \emph{IEEE Transactions on Very Large Scale Integration (VLSI)
  Systems}, vol.~28, no.~9, pp. 1953--1965, 2020.

\bibitem{moreno2020analysis}
A.~A. Moreno, J.~Olivito, J.~Resano, and H.~Mecha, ``{Analysis of a Pipelined
  Architecture for Sparse DNNs on Embedded Systems},'' \emph{IEEE Transactions
  on Very Large Scale Integration (VLSI) Systems}, vol.~28, no.~9, pp.
  1993--2003, 2020.

\bibitem{lian2019high}
X.~Lian, Z.~Liu, Z.~Song, J.~Dai, W.~Zhou, and X.~Ji, ``{High-Performance
  FPGA-based CNN Accelerator with Block-Floating-Point Arithmetic},''
  \emph{IEEE Transactions on Very Large Scale Integration (VLSI) Systems},
  vol.~27, no.~8, pp. 1874--1885, 2019.

\bibitem{kala2019high}
S.~Kala, B.~R. Jose, J.~Mathew, and S.~Nalesh, ``{High-Performance CNN
  Accelerator on FPGA using Unified Winograd-GEMM Architecture},'' \emph{IEEE
  Transactions on Very Large Scale Integration (VLSI) Systems}, vol.~27,
  no.~12, pp. 2816--2828, 2019.

\bibitem{xie2021efficient}
X.~Xie, J.~Lin, Z.~Wang, and J.~Wei, ``{An Efficient and Flexible Accelerator
  Design for Sparse Convolutional Neural Networks},'' \emph{IEEE Transactions
  on Circuits and Systems I: Regular Papers (TCAS-I)}, vol.~68, no.~7, pp.
  2936--2949, 2021.

\bibitem{albericio2016cnvlutin}
J.~Albericio, P.~Judd, T.~Hetherington, T.~Aamodt, N.~E. Jerger, and
  A.~Moshovos, ``{Cnvlutin: Ineffectual-Neuron-Free Deep Neural Network
  Computing},'' in \emph{2016 ACM/IEEE 43rd Annual International Symposium on
  Computer Architecture (ISCA)}.\hskip 1em plus 0.5em minus 0.4em\relax IEEE,
  2016, pp. 1--13.

\bibitem{chen2016eyeriss}
Y.-H. Chen, J.~Emer, and V.~Sze, ``{Eyeriss: A Spatial Architecture for
  Energy-Efficient Dataflow for Convolutional Neural Networks},'' in \emph{2016
  ACM/IEEE 43rd Annual International Symposium on Computer Architecture
  (ISCA)}.\hskip 1em plus 0.5em minus 0.4em\relax IEEE, 2016, pp. 367--379.

\bibitem{liu2016cambricon}
S.~Liu, Z.~Du, J.~Tao, D.~Han, T.~Luo, Y.~Xie, Y.~Chen, and T.~Chen,
  ``{Cambricon: An Instruction Set Architecture for Neural Networks},'' in
  \emph{2016 ACM/IEEE 43rd Annual International Symposium on Computer
  Architecture (ISCA)}.\hskip 1em plus 0.5em minus 0.4em\relax IEEE, 2016, pp.
  393--405.

\bibitem{liu2020systolic}
Z.-G. Liu, P.~N. Whatmough, and M.~Mattina, ``{Systolic Tensor Array: An
  Efficient Structured-sparse GEMM Accelerator for Mobile CNN Inference},''
  \emph{IEEE Computer Architecture Letters}, vol.~19, no.~1, pp. 34--37, 2020.

\bibitem{parashar2017scnn}
A.~Parashar, M.~Rhu, A.~Mukkara, A.~Puglielli, R.~Venkatesan, B.~Khailany,
  J.~Emer, S.~W. Keckler, and W.~J. Dally, ``{SCNN: An Accelerator for
  Compressed-sparse Convolutional Neural Networks},'' in \emph{Proceedings of
  the 44th Annual International Symposium on Computer Architecture (ISCA)},
  2017, pp. 27--40.

\bibitem{li2020ftrans}
B.~Li, S.~Pandey, H.~Fang, Y.~Lyv, J.~Li, J.~Chen, M.~Xie, L.~Wan, H.~Liu, and
  C.~Ding, ``{FTRANS: Energy-Efficient Acceleration of Transformers using
  FPGA},'' in \emph{Proceedings of the ACM/IEEE International Symposium on Low
  Power Electronics and Design (ISLPED)}, 2020.

\bibitem{ham2020a3}
T.~J. Ham, S.~J. Jung, S.~Kim, Y.~H. Oh, Y.~Park, Y.~Song, J.-H. Park, S.~Lee,
  K.~Park, J.~W. Lee \emph{et~al.}, ``{A\textsuperscript{3}: Accelerating
  Attention Mechanisms in Neural Networks with Approximation},'' in \emph{2020
  IEEE International Symposium on High Performance Computer Architecture
  (HPCA)}, 2020.

\bibitem{wang2021spatten}
H.~Wang, Z.~Zhang, and S.~Han, ``{SpAtten: Efficient Sparse Attention
  Architecture with Cascade Token and Head Pruning},'' in \emph{2021 IEEE
  International Symposium on High-Performance Computer Architecture (HPCA)},
  2021.

\bibitem{lu2021sanger}
L.~Lu, Y.~Jin, H.~Bi, Z.~Luo, P.~Li, T.~Wang, and Y.~Liang, ``{Sanger: A
  Co-Design Framework for Enabling Sparse Attention using Reconfigurable
  Architecture},'' in \emph{Proceedings of the 54th Annual IEEE/ACM
  International Symposium on Microarchitecture (MICRO)}, 2021.

\bibitem{peng2021accelerating}
H.~Peng, S.~Huang, T.~Geng, A.~Li, W.~Jiang, H.~Liu, S.~Wang, and C.~Ding,
  ``{Accelerating Transformer-based Deep Learning Models on FPGAs using Column
  Balanced Block Pruning},'' in \emph{2021 22nd International Symposium on
  Quality Electronic Design (ISQED)}, 2021.

\bibitem{lin2020dynamic}
T.~Lin, S.~U. Stich, L.~Barba, D.~Dmitriev, and M.~Jaggi, ``Dynamic model
  pruning with feedback,'' \emph{arXiv preprint arXiv:2006.07253}, 2020.

\bibitem{hoefler2021sparsity}
T.~Hoefler, D.~Alistarh, T.~Ben-Nun, N.~Dryden, and A.~Peste, ``Sparsity in
  deep learning: Pruning and growth for efficient inference and training in
  neural networks,'' \emph{Journal of Machine Learning Research (JMLR)},
  vol.~22, no. 241, pp. 1--124, 2021.

\bibitem{zhang2016cambricon}
S.~Zhang, Z.~Du, L.~Zhang, H.~Lan, S.~Liu, L.~Li, Q.~Guo, T.~Chen, and Y.~Chen,
  ``{Cambricon-X: An accelerator for sparse neural networks},'' in \emph{2016
  49th Annual IEEE/ACM International Symposium on Microarchitecture
  (MICRO)}.\hskip 1em plus 0.5em minus 0.4em\relax IEEE, 2016, pp. 1--12.

\bibitem{wang2018glue}
A.~Wang, A.~Singh, J.~Michael, F.~Hill, O.~Levy, and S.~Bowman, ``{GLUE: A
  Multi-Task Benchmark and Analysis Platform for Natural Language
  Understanding},'' in \emph{Proceedings of the 2018 EMNLP Workshop
  BlackboxNLP: Analyzing and Interpreting Neural Networks for NLP}, 2018, pp.
  353--355.

\bibitem{jiao2020tinybert}
X.~Jiao, Y.~Yin, L.~Shang, X.~Jiang, X.~Chen, L.~Li, F.~Wang, and Q.~Liu,
  ``{TinyBERT: Distilling BERT for Natural Language Understanding},'' in
  \emph{Proceedings of the 2020 Conference on Empirical Methods in Natural
  Language Processing (EMNLP): Findings}, 2020, pp. 4163--4174.

\bibitem{caron2021emerging}
M.~Caron, H.~Touvron, I.~Misra, H.~J\'egou, J.~Mairal, P.~Bojanowski, and
  A.~Joulin, ``{Emerging Properties in Self-Supervised Vision Transformers},''
  in \emph{Proceedings of the International Conference on Computer Vision
  (ICCV)}, 2021.

\bibitem{wolf2019huggingface}
T.~Wolf, L.~Debut, V.~Sanh, J.~Chaumond, C.~Delangue, A.~Moi, P.~Cistac,
  T.~Rault, R.~Louf, M.~Funtowicz \emph{et~al.}, ``{Huggingface's Transformers:
  State-of-the-art Natural Language Processing},'' \emph{arXiv preprint
  arXiv:1910.03771}, 2019.

\bibitem{taylor2018basejump}
M.~B. Taylor, ``{Basejump STL: SystemVerilog Needs a Standard Template Library
  for Hardware Design},'' in \emph{2018 55th ACM/ESDA/IEEE Design Automation
  Conference (DAC)}.\hskip 1em plus 0.5em minus 0.4em\relax IEEE, 2018, pp.
  1--6.

\bibitem{rossi2015pulp}
D.~Rossi, F.~Conti, A.~Marongiu, A.~Pullini, I.~Loi, M.~Gautschi,
  G.~Tagliavini, A.~Capotondi, P.~Flatresse, and L.~Benini, ``{PULP: A Parallel
  Ultra Low Power Platform for Next Generation IoT Applications},'' in
  \emph{2015 IEEE Hot Chips 27 Symposium (HCS)}.\hskip 1em plus 0.5em minus
  0.4em\relax IEEE, 2015, pp. 1--39.

\end{thebibliography}
%



\end{document}